\pdfoutput=1
\documentclass[12pt,a4paper]{article}

\usepackage{subcaption}
\usepackage{ifthen} 
\newboolean{pdflatex}
\setboolean{pdflatex}{true} 

\newboolean{articletitles}
\setboolean{articletitles}{true} 

\newboolean{uprightparticles}
\setboolean{uprightparticles}{false} 

\def\paperauthors{LHCb collaboration} 
\def\paperasciititle{Measurement of Jpsi-pair production cross-sections in pp collisions at sqrt{s}=13 TeV and study of gluon TMDs} 
\def\papertitle{Measurement of \jpsi-pair production \\ in $pp$ collisions at $\sqs=13\tev$ and \\ \mbox{study of gluon transverse-momentum} \\ dependent PDFs} 
\def\paperkeywords{{High Energy Physics}, {LHCb}} 
\def\papercopyright{\the\year\ CERN for the benefit of the LHCb collaboration} 
\def\paperlicence{CC BY 4.0 licence}
\def\paperlicenceurl{https://creativecommons.org/licenses/by/4.0/}

\usepackage[top=1in, bottom=1.25in, left=1in, right=1in]{geometry}

\usepackage{microtype}
\usepackage{lineno}  
\usepackage{xspace} 
\usepackage{caption} 

\usepackage{graphicx}  
\usepackage{color}
\usepackage{colortbl}
\graphicspath{{./figs/}} 

\usepackage{amsmath} 
\usepackage{amssymb}
\usepackage{amsfonts}
\usepackage{upgreek} 

\newcommand*\patchAmsMathEnvironmentForLineno[1]{%
\expandafter\let\csname old#1\expandafter\endcsname\csname #1\endcsname
\expandafter\let\csname oldend#1\expandafter\endcsname\csname
end#1\endcsname
 \renewenvironment{#1}%
   {\linenomath\csname old#1\endcsname}%
   {\csname oldend#1\endcsname\endlinenomath}%
}
\newcommand*\patchBothAmsMathEnvironmentsForLineno[1]{%
  \patchAmsMathEnvironmentForLineno{#1}%
  \patchAmsMathEnvironmentForLineno{#1*}%
}
\AtBeginDocument{%
\patchBothAmsMathEnvironmentsForLineno{equation}%
\patchBothAmsMathEnvironmentsForLineno{align}%
\patchBothAmsMathEnvironmentsForLineno{flalign}%
\patchBothAmsMathEnvironmentsForLineno{alignat}%
\patchBothAmsMathEnvironmentsForLineno{gather}%
\patchBothAmsMathEnvironmentsForLineno{multline}%
\patchBothAmsMathEnvironmentsForLineno{eqnarray}%
}

\usepackage{hyperxmp}

\usepackage[pdftex,
            pdfauthor={\paperauthors},
            pdftitle={\paperasciititle},
            pdfkeywords={\paperkeywords},
            pdfcopyright={Copyright (C) \papercopyright},
            pdflicenseurl={\paperlicenceurl}]{hyperref}

\usepackage[colorinlistoftodos,textsize=scriptsize]{todonotes}

\usepackage[bottom,flushmargin,hang,multiple]{footmisc}

\usepackage[all]{hypcap} 

\usepackage{xspace} 
\usepackage{upgreek}

\def\lhcb   {\mbox{LHCb}\xspace}

\def\MagUp {\mbox{\em Mag\kern -0.05em Up}\xspace}

\ifthenelse{\boolean{uprightparticles}}%
{

 \def\Pmu         {\ensuremath{\upmu}\xspace}

 \def\Ppsi        {\ensuremath{\uppsi}\xspace}

 \def\PDelta      {\ensuremath{\Delta}\xspace}                 
 \def\PXi         {\ensuremath{\Xi}\xspace}                 
 \def\PLambda     {\ensuremath{\Lambda}\xspace}                 
 \def\PSigma      {\ensuremath{\Sigma}\xspace}                 
 \def\POmega      {\ensuremath{\Omega}\xspace}                 
 \def\PUpsilon    {\ensuremath{\Upsilon}\xspace}
 \let\oldPi\Pi
 \def\PPi         {\ensuremath{\oldPi}\xspace}

 \def\PB      {\ensuremath{\mathrm{B}}\xspace}                 
                  
 \def\PD      {\ensuremath{\mathrm{D}}\xspace}

 \def\PJ      {\ensuremath{\mathrm{J}}\xspace}                 
 \def\PK      {\ensuremath{\mathrm{K}}\xspace}

 \def\Pb      {\ensuremath{\mathrm{b}}\xspace}                 
 \def\Pc      {\ensuremath{\mathrm{c}}\xspace}

 \def\Pi      {\ensuremath{\mathrm{i}}\xspace}

 \def\Pp      {\ensuremath{\mathrm{p}}\xspace}

 \def\Ps      {\ensuremath{\mathrm{s}}\xspace}

 \def\thebaroffset{0.0em}
}
{

 \def\Pmu         {\ensuremath{\mu}\xspace}

 \def\Ppsi        {\ensuremath{\psi}\xspace}                 
                  
 \mathchardef\PDelta="7101
 \mathchardef\PXi="7104
 \mathchardef\PLambda="7103
 \mathchardef\PSigma="7106
 \mathchardef\POmega="710A
 \mathchardef\PUpsilon="7107
 \mathchardef\PPi="7105
                  
 \def\PB      {\ensuremath{B}\xspace}                 
                  
 \def\PD      {\ensuremath{D}\xspace}

 \def\PJ      {\ensuremath{J}\xspace}                 
 \def\PK      {\ensuremath{K}\xspace}

 \def\Pb      {\ensuremath{b}\xspace}                 
 \def\Pc      {\ensuremath{c}\xspace}

 \def\Pi      {\ensuremath{i}\xspace}

 \def\Pp      {\ensuremath{p}\xspace}

 \def\Ps      {\ensuremath{s}\xspace}

 \def\thebaroffset{0.18em}
}
\newcommand{\offsetoverline}[2][\thebaroffset]{\kern #1\overline{\kern -#1 #2}}%

\makeatletter
\ifcase \@ptsize \relax
  \newcommand{\miniscule}{\@setfontsize\miniscule{4}{5}}
\or
  \newcommand{\miniscule}{\@setfontsize\miniscule{5}{6}}
\or
  \newcommand{\miniscule}{\@setfontsize\miniscule{5}{6}}
\fi
\makeatother

\DeclareRobustCommand{\optbar}[1]{\shortstack{{\miniscule (\rule[.5ex]{1.25em}{.18mm})}
  \\ [-.7ex] $#1$}}

\def\mumu       {{\ensuremath{\Pmu^+\Pmu^-}}\xspace}

\def\squark    {{\ensuremath{\Ps}}\xspace}

\def\cquark    {{\ensuremath{\Pc}}\xspace}

\def\bquark    {{\ensuremath{\Pb}}\xspace}

\def\KorKbar {\kern \thebaroffset\optbar{\kern -\thebaroffset \PK}{}\xspace}

\def\D       {{\ensuremath{\PD}}\xspace}

\def\DorDbar {\kern \thebaroffset\optbar{\kern -\thebaroffset \PD}\xspace}

\def\Dp      {{\ensuremath{\D^+}}\xspace}
\def\Dm      {{\ensuremath{\D^-}}\xspace}

\def\DpDm    {\ensuremath{\Dp {\kern -0.16em \Dm}}\xspace}

\def\B       {{\ensuremath{\PB}}\xspace}

\def\BorBbar {\kern \thebaroffset\optbar{\kern -\thebaroffset \PB}\xspace}

\def\Bd      {{\ensuremath{\B^0}}\xspace}

\def\BdorBdbar {\kern \thebaroffset\optbar{\kern -\thebaroffset \Bd}\xspace}

\def\Bs      {{\ensuremath{\B^0_\squark}}\xspace}

\def\BsorBsbar {\kern \thebaroffset\optbar{\kern -\thebaroffset \Bs}\xspace}

\def\jpsi     {{\ensuremath{{\PJ\mskip -3mu/\mskip -2mu\Ppsi}}}\xspace}

\def\Y#1S{\ensuremath{\PUpsilon{(#1S)}}\xspace}
\def\OneS  {{\Y1S}\xspace}

\def\proton      {{\ensuremath{\Pp}}\xspace}
\def\antiproton  {{\ensuremath{\overline \proton}}\xspace}

\def\LorLbar     {\kern \thebaroffset\optbar{\kern -\thebaroffset \PLambda}\xspace}

\def\BF         {{\ensuremath{\mathcal{B}}}\xspace}
\def\BR         {\BF}

\def\to                 {\ensuremath{\rightarrow}\xspace}

\def\eps   {{\ensuremath{\varepsilon}}\xspace}

\newcommand{\etot}{\ensuremath{\varepsilon^{\mathrm{ tot}}}\xspace}
\def\eacc {\ensuremath{\eps^{\rm acc}}\xspace}
\def\erecsel {\ensuremath{\eps^{\rm rec\&sel}}\xspace}
\def\etri {\ensuremath{\eps^{\rm trig}}\xspace}
\def\epid {\ensuremath{\eps^{\rm PID}}\xspace}

\def\AT#1     {\ensuremath{A_{\mathrm{T}}^{#1}}\xspace}           

\def\C#1      {\ensuremath{\mathcal{C}_{#1}}\xspace}                       
\def\Cp#1     {\ensuremath{\mathcal{C}_{#1}^{'}}\xspace}                    
\def\Ceff#1   {\ensuremath{\mathcal{C}_{#1}^{\mathrm{(eff)}}}\xspace}        
\def\Cpeff#1  {\ensuremath{\mathcal{C}_{#1}^{'\mathrm{(eff)}}}\xspace}       
\def\Ope#1    {\ensuremath{\mathcal{O}_{#1}}\xspace}                       
\def\Opep#1   {\ensuremath{\mathcal{O}_{#1}^{'}}\xspace}                    

\newcommand{\nospaceunit}[1]{\ensuremath{\text{#1}}}       
\newcommand{\aunit}[1]{\ensuremath{\text{\,#1}}}       

\newcommand{\tev}{\aunit{Te\kern -0.1em V}\xspace}
\newcommand{\gev}{\aunit{Ge\kern -0.1em V}\xspace}
\newcommand{\mev}{\aunit{Me\kern -0.1em V}\xspace}
\newcommand{\kev}{\aunit{ke\kern -0.1em V}\xspace}
\newcommand{\ev}{\aunit{e\kern -0.1em V}\xspace}
 
\newcommand{\mevc}{\ensuremath{\aunit{Me\kern -0.1em V\!/}c}\xspace}
\newcommand{\gevc}{\ensuremath{\aunit{Ge\kern -0.1em V\!/}c}\xspace}
\newcommand{\mevcc}{\ensuremath{\aunit{Me\kern -0.1em V\!/}c^2}\xspace}
\newcommand{\gevcc}{\ensuremath{\aunit{Ge\kern -0.1em V\!/}c^2}\xspace}

\def\mum  {\ensuremath{\,\upmu\nospaceunit{m}}\xspace}

\def\mbarn{\aunit{mb}\xspace}
\def\mub{\ensuremath{\,\upmu\nospaceunit{b}}\xspace}
\def\nb {\aunit{nb}\xspace}

\def\fb   {\ensuremath{\aunit{fb}}\xspace}
\def\invfb   {\ensuremath{\fb^{-1}}\xspace}

\def\ps   {\ensuremath{\aunit{ps}}\xspace}

\newcommand{\chisq}{\ensuremath{\chi^2}\xspace}

\newcommand{\chisqip}{\ensuremath{\chi^2_{\text{IP}}}\xspace}

\def\deriv {\ensuremath{\mathrm{d}}}

\def\gsim{{~\raise.15em\hbox{$>$}\kern-.85em
          \lower.35em\hbox{$\sim$}~}\xspace}
\def\lsim{{~\raise.15em\hbox{$<$}\kern-.85em
          \lower.35em\hbox{$\sim$}~}\xspace}

\def\sPlot{\mbox{\em sPlot}\xspace}

\def\sqs   {\ensuremath{\protect\sqrt{s}}\xspace}

\def\pt         {\ensuremath{p_{\mathrm{T}}}\xspace}

\def\ptot       {\ensuremath{p}\xspace}

\newcommand{\lum} {\ensuremath{\mathcal{L}}\xspace}

\def\evtgen     {\mbox{\textsc{EvtGen}}\xspace}

\def\geant      {\mbox{\textsc{Geant4}}\xspace}

\def\photos     {\mbox{\textsc{Photos}}\xspace}

\def\pythia     {\mbox{\textsc{Pythia}}\xspace}

\def\tell1  {TELL1\xspace}
\def\ukl1   {UKL1\xspace}

\newcommand{\ie}{\mbox{\itshape i.e.}\xspace}

\newcommand{\lhcborcid}[1]{\href{https://orcid.org/#1}{\hspace*{0.1em}\raisebox{-0.45ex}{\includegraphics[width=1em]{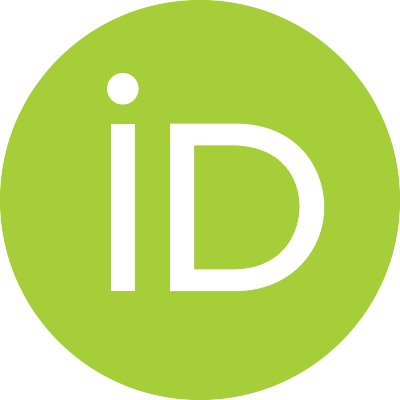}}}}

\usepackage{cite} 
\usepackage{mciteplus}

\usepackage{longtable} 

\begin{document}

\renewcommand{\thefootnote}{\fnsymbol{footnote}}
\setcounter{footnote}{1}


\begin{titlepage}
\pagenumbering{roman}

\vspace*{-1.5cm}
\centerline{\large EUROPEAN ORGANIZATION FOR NUCLEAR RESEARCH (CERN)}
\vspace*{1.5cm}
\noindent
\begin{tabular*}{\linewidth}{lc@{\extracolsep{\fill}}r@{\extracolsep{0pt}}}
\ifthenelse{\boolean{pdflatex}}
{\vspace*{-1.5cm}\mbox{\!\!\!\includegraphics[width=.14\textwidth]{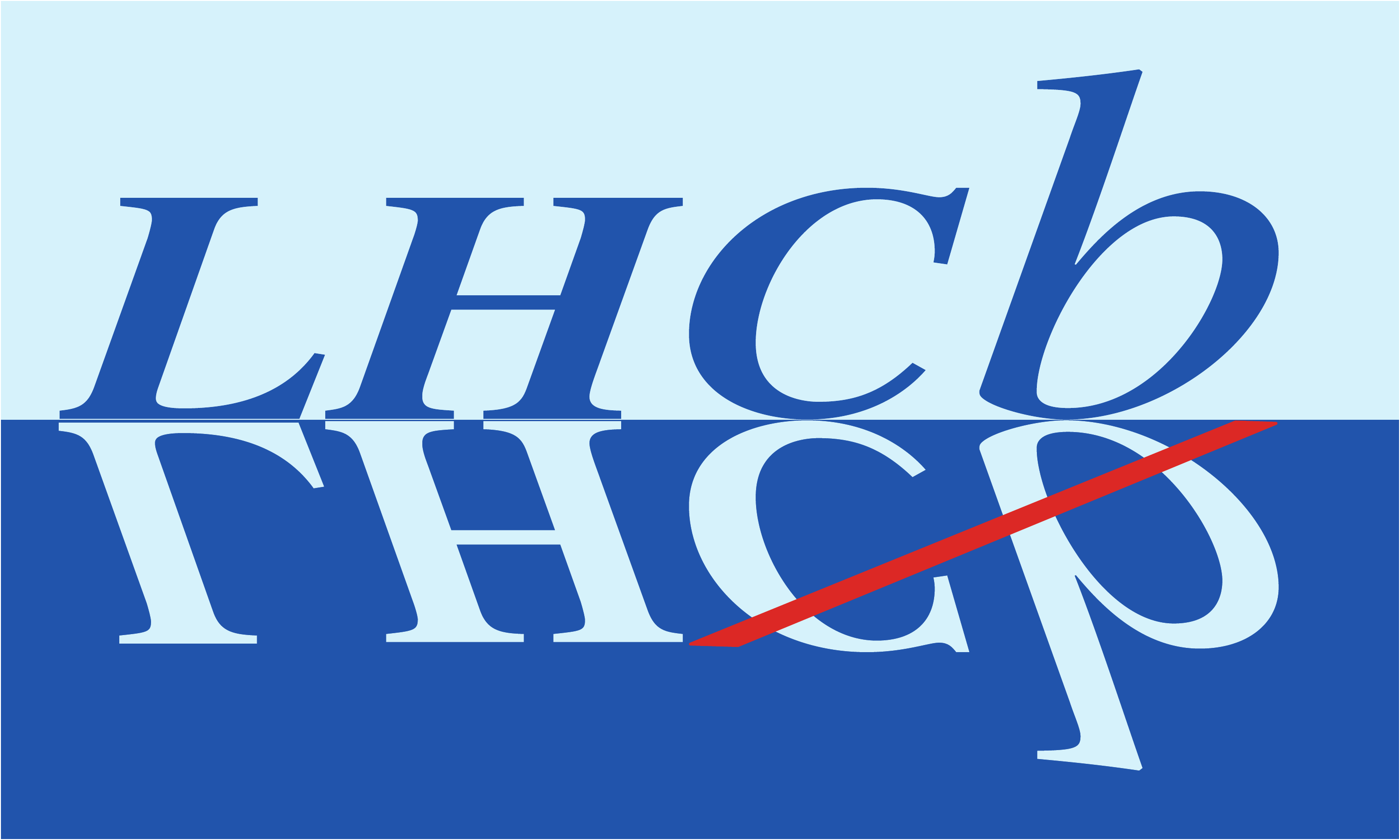}} & &}%
{\vspace*{-1.2cm}\mbox{\!\!\!\includegraphics[width=.12\textwidth]{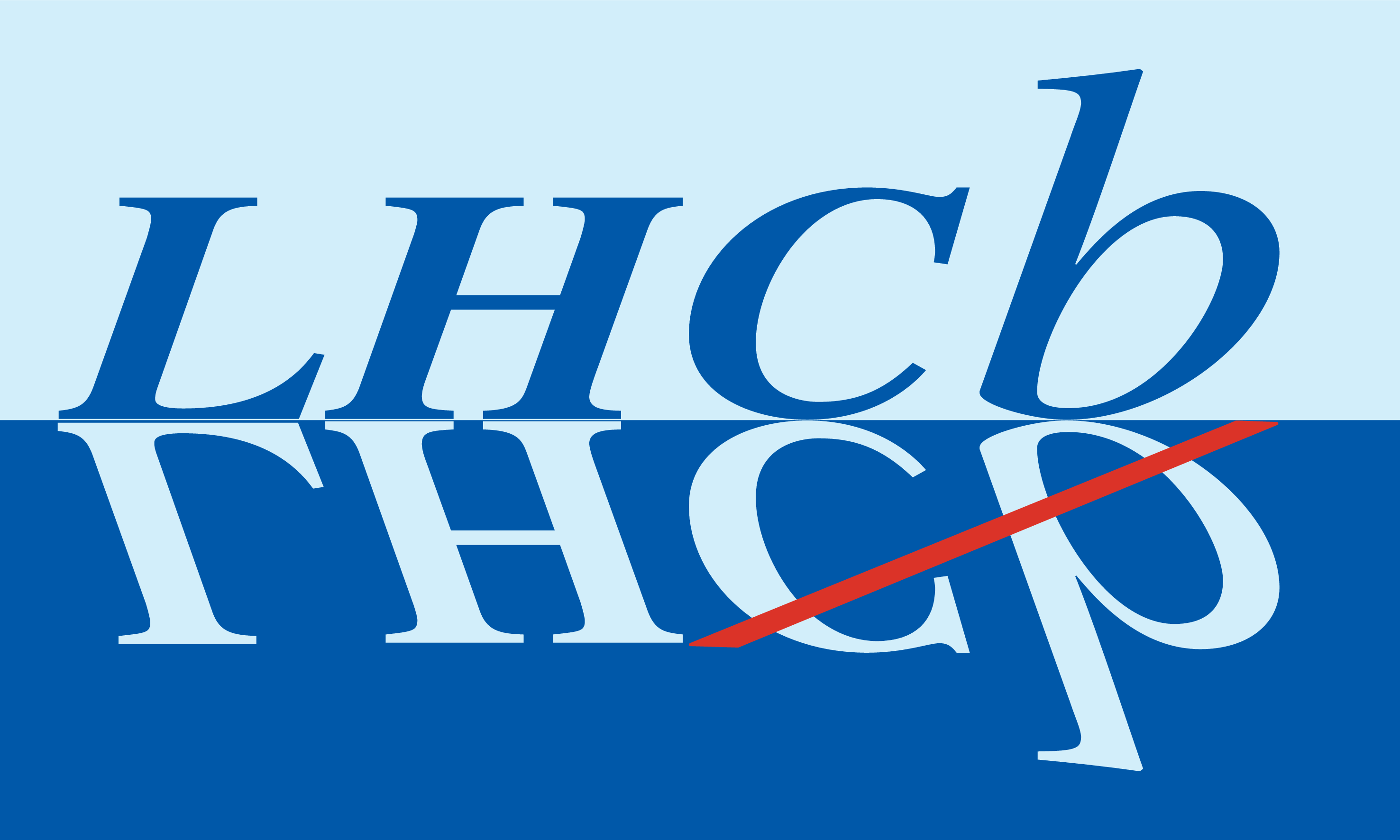}} & &}%
\\
 & & CERN-EP-2023-242 \\  
 & & LHCb-PAPER-2023-022 \\  
 & & March 14, 2024 \\ 
 & & \\
\end{tabular*}

\vspace*{4.0cm}

{\normalfont\bfseries\boldmath\huge
\begin{center}
  \papertitle 
\end{center}
}

\vspace*{1.0cm}

\begin{center}
\paperauthors\footnote{Authors are listed at the end of this paper.}
\end{center}

\vspace{\fill}

\begin{abstract}
  \noindent
  The production cross-section of \jpsi pairs in proton-proton collisions at a centre-of-mass energy of $\sqs=13\tev$ is measured using a data sample corresponding to an integrated luminosity of $4.2\invfb$ collected by the LHCb experiment.
  The measurement is performed with both \jpsi mesons in the transverse momentum range $0<\pt<14\gevc$ and rapidity range $2.0<y<4.5$.
  The cross-section of this process is measured to be $16.36\pm0.28~(\text{stat})\pm0.88~(\text{syst})\nb$.
  The contributions from single-parton scattering and double-parton scattering are separated based on the dependence of the cross-section on the absolute rapidity difference $\Delta y$ between the two \jpsi mesons.
  The effective cross-section of double-parton scattering is measured to be $\sigma_{\text{eff}}=13.1\pm1.8~(\text{stat})\pm2.3~(\text{syst})\mbarn$.
  The distribution of the azimuthal angle $\phi_{\text{CS}}$ of one of the \jpsi mesons in the Collins-Soper frame and the \pt-spectrum of the \jpsi pairs are also measured for the study of the gluon transverse-momentum dependent distributions inside protons.
  The extracted values of $\langle\cos2\phi_{\text{CS}}\rangle$ and $\langle\cos4\phi_{\text{CS}}\rangle$ are consistent with zero, but the presence of azimuthal asymmetry at a few percent level is allowed.
  
\end{abstract}

\vspace*{0.5cm}

\begin{center}
  Published in JHEP 03 (2024) 088 
\end{center}

\vspace{\fill}

{\footnotesize 
\centerline{\copyright~\papercopyright. \href{\paperlicenceurl}{\paperlicence}.}}
\vspace*{2mm}

\end{titlepage}


\newpage
\setcounter{page}{2}
\mbox{~}
%
%
%
%


\renewcommand{\thefootnote}{\arabic{footnote}}
\setcounter{footnote}{0}

\cleardoublepage


\pagestyle{plain} 
\setcounter{page}{1}
\pagenumbering{arabic}


\section{Introduction}
\label{sec:Introduction}
In the Standard Model~(SM) of particle physics, the strong interaction is described by quantum chromodynamics~(QCD).
In the low-energy regime, the QCD coupling constant evolves to be so large that perturbation theory is not valid any more.
The details of how the fundamental quarks and gluons are distributed inside hadrons and dynamically generate the hadron mass and spin are still largely unknown.
As a building block of the physical world,
the proton lies at the forefront of hadron structure studies.
Moreover, the knowledge of the partonic structure of the proton is an essential input to the majority of measurements at hadron colliders, including searches for physics beyond the SM,
among which the $W$-mass measurement is a typical example~\cite{CDF:2022hxs,D0:2012kms,ATLAS:2017rzl,LHCb:2021bjt}.

In the past, the internal structure of the proton was mainly studied in terms of one-dimensional parton distribution functions~(PDFs)
that parameterise the longitudinal momentum fraction~(usually denoted $x$) distributions of partons inside the proton.
Recently, significant progress has been made in constructing the theoretical framework for transverse-momentum dependent parton distribution functions~(TMDs)~\cite{Mulders:2000sh,Dominguez:2011wm,Echevarria:2012js},
leading to a more comprehensive understanding of the proton structure.
The quark TMDs have been studied through semi-inclusive deep-inelastic scattering and Drell-Yan measurements at HERMES, COMPASS and E866/NuSea experiments, and at a series of experiments at JLab~\cite{HERMES:2009lmz,COMPASS:2023cgk,NuSea:2006gvb,Barry:2023qqh,JeffersonLabHallA:2013qky,CLAS:2010fns}.
The gluon TMDs, however, are much less known,
so probing gluon TMDs is one of the major objectives of future experimental facilities like the Electron Ion Collider~(EIC)~\cite{Boer:2011fh}, LHCSpin~\cite{Passalacqua:2022jia} and LHC fixed-target experiments~\cite{Brodsky:2012vg}.
In unpolarised protons, the gluon TMDs can be parameterised at leading twist using two TMDs~\cite{Mulders:2000sh}:
the distribution of unpolarised gluons $f_{1}^{g}(x,k_{\rm T}^{2},\mu)$ and that of linearly polarised gluons $h_{1}^{\perp g}(x,k_{\rm T}^{2},\mu)$,
in which $k_{\rm T}$ is the gluon transverse momentum
and $\mu$ is the factorisation scale.
In particular, the knowledge on the $h_{1}^{\perp g}(x,k_{\rm T}^{2},\mu)$ function is still very limited.
The production of $\jpsi$ pairs in proton-proton~($pp$) collisions through single-parton scattering~(SPS) has been proposed as the golden channel to probe gluon TMDs,
in which the presence of linearly polarised gluons will lead to azimuthal asymmetries at the percent level~\cite{Lansberg:2017dzg,Scarpa:2019fol}.
The transverse momentum~($\pt$) spectrum of $\jpsi$ pairs also encodes information on $f_{1}^{g}(x,k_{\rm T}^{2},\mu)$.
In fact, the differential production cross-section of $\jpsi$ pairs as a function of $\pt$ measured by the \lhcb experiment using the 2015 data was used to perform the first fit of $f_{1}^{g}(x,k_{\rm T}^{2},\mu)$ and obtain $\langle k_{\rm T}^{2}\rangle$ at an effective factorisation scale~\cite{LHCb-PAPER-2016-057,Lansberg:2017dzg}.

Quarkonium production is also one of the best tools to study hadronisation.
The non-relativistic QCD~(NRQCD) model provides the most successful description of quarkonium production so far,
but it still can not describe coherently the production and polarisation of various quarkonium states measured in different collisions~\cite{Bodwin:1994jh,Cho:1995vh,Cho:1995ce,Zhang:2020atv}.
The SPS production of $\jpsi$ pairs~(also referred to as di-\jpsi hereafter) can add valuable information to solve this puzzle~\cite{Sun:2014gca,Lansberg:2019fgm,Likhoded:2016zmk,PhysRevD.84.054012}.
In addition to SPS, quarkonium pairs can be produced through double-parton scattering~(DPS)~\cite{Lansberg:2014swa}.
It is a process of great interest, which has been widely studied by many experiments via various reactions~\cite{LHCb:2023qgu,ATLAS:2014ofp,ATLAS:2016ydt,LHCb:2015wvu,ATLAS:2014yjd,CMS:2014cmt,LHCb:2012aiv,ATLAS:2013aph,CMS:2013huw,D0:2015dyx,D0:2014vql,D0:2009apj,CDF:1993sbj,CDF:1997yfa}.
It can be used, for instance, to reveal the profile and correlation of partons inside the proton,
which are encoded in a characteristic parameter of DPS called effective cross-section, denoted as $\sigma_{\rm eff}$~\cite{Calucci:1997ii, Calucci:1999yz, DelFabbro:2000ds}.
The contribution from DPS can be estimated according to the formula~\cite{Calucci:1997ii, Calucci:1999yz, DelFabbro:2000ds}
\begin{equation}
\label{eq:pocket}
    \sigma^{\text{DPS}}_{\text{di-}\jpsi}= \frac{1}{2}\frac{\sigma^2_{\jpsi}}{\sigma_{\text{eff}}},
\end{equation}
where $\sigma_{\jpsi}$ is the prompt \jpsi meson production cross-section, and the factor one-half accounts for the two identical particles in the final state.
The effective cross-section $\sigma_{\text{eff}}$ characterises the transverse overlap area between the interacting partons.

In this paper, the $\jpsi$-pair production cross-section in $pp$ collisions at a centre-of-mass energy of ${\sqs=13\tev}$ is measured for $\jpsi$ mesons with ${\pt<14\gevc}$ and rapidity ${2.0<y<4.5}$
using a subset of data collected by the \lhcb experiment from 2016 to 2018 with specific trigger requirements,
corresponding to an integrated luminosity of ${4.2\invfb}$.
The azimuthal asymmetry of $\jpsi$ pairs is measured to probe the TMD function $h_{1}^{\perp g}(x,k_{\rm T}^{2},\mu)$,
presenting the first experimental measurement of linear polarisation of gluons inside unpolarised protons.
The \pt spectrum of the \jpsi pairs is measured in intervals of \jpsi-pair rapidity and mass,
which will help to extract $f_{1}^{g}(x,k_{\rm T}^{2},\mu)$ with the TMD evolution effect considered~\cite{Scarpa:2019fol}.
Updates of the differential production cross-sections given in the previous LHCb measurement using data with a luminosity of about 0.3\invfb at ${\sqs=13\tev}$~\cite{LHCb-PAPER-2016-057} are also provided,
with the SPS and DPS contributions separated without dependence on any specific SPS production model.

\section{Detector and simulation}
\label{sec:Detector}
The \lhcb detector~\cite{LHCb-DP-2008-001,LHCb-DP-2014-002} is a single-arm forward spectrometer covering the \mbox{pseudorapidity} range $2<\eta <5$, designed for the study of particles containing \bquark- or \cquark-quarks.
The detector includes a high-precision tracking system consisting of a silicon-strip vertex detector surrounding the $pp$ interaction region, a large-area silicon-strip detector located upstream of a dipole magnet with a bending power of about $4{\mathrm{\,Tm}}$, and three stations of silicon-strip detectors and straw drift tubes placed downstream of the magnet.
The tracking system provides a measurement of the momentum, \ptot, of charged particles with a relative uncertainty that varies from 0.5\% at low momentum to 1.0\% at 200\gevc.
The minimum distance of a track to a primary $pp$ collision vertex (PV), the impact parameter (IP), is measured with a resolution of $(15+29/\pt)\mum$, where \pt is in\,\gevc.
Different types of charged hadrons are distinguished using information from two ring-imaging Cherenkov detectors. 
Photons, electrons and hadrons are identified by a calorimeter system consisting of scintillating-pad and preshower detectors, an electromagnetic and a hadronic calorimeter.
Muons are identified by a system composed of alternating layers of iron and multiwire proportional chambers.

Simulated samples of \jpsi mesons are produced to study the expected behaviour of experimental signals and determine the detection efficiencies.
The $pp$ collisions are modelled using \pythia~\cite{Sjostrand:2007gs,Sjostrand:2006za} with a specific \lhcb configuration~\cite{LHCb-PROC-2010-056}.
In the \pythia model, \jpsi mesons are not polarised, and the leading order colour-singlet and colour-octet contributions~\cite{LHCb-PROC-2010-056,Bargiotti:2007zz} are included in prompt \jpsi meson production.
Decays of unstable particles are described by \evtgen~\cite{Lange:2001uf} with QED final-state radiation handled by \photos~\cite{davidson2015photos}.
The interactions of the generated particles with the detector are modelled using the \geant toolkit~\cite{Allison:2006ve,Agostinelli:2002hh} as described in Ref.~\cite{LHCb-PROC-2011-006}.

\section{Candidate selection}
\label{sec:Selection}
The di-\jpsi candidates are reconstructed through the $\jpsi \to \mumu$ decays.
The online event selection is performed by a trigger~\cite{LHCb-DP-2012-004}, 
which consists of a hardware stage~(L0), based on information from the calorimeter and muon
systems, followed by a two-step software stage~(HLT1 and HLT2), which applies a full event
reconstruction.
At least one $\jpsi$ meson is required to fulfil the selection criteria of the L0 and HLT1 triggers.
The L0 trigger selects two muons with the product of their transverse momenta larger than $1.3^2$, $1.5^2$ or $1.8^2\,(\text{Ge\kern -0.1em V}/c)^2$, depending on the data taking period.
The HLT1 trigger requires two good-quality tracks with ${\pt>0.3\gevc}$ and ${p>6\gevc}$, that are loosely identified as muons and form a \jpsi candidate with an invariant mass ${m_{\mumu}>2.7\gevcc}$ or 2.9\gevcc, depending on the data taking period.
The HLT2 trigger requires both $\mumu$ pairs to form a good vertex and have an invariant mass $m_{\mumu}$ within ${120\mevcc}$ of the world-average value of the \jpsi mass~\cite{Workman:2022ynf}.
Offline selections are applied to the di-\jpsi candidates to further reduce the combinatorial background.
All the four muon tracks are required to have ${1.9<\eta<4.9}$, ${\pt>0.65\gevc}$ and ${p>3\gevc}$.

In $pp$ collisions, \jpsi mesons can be produced promptly at the PV, or in the decays of beauty hadrons.
The nonprompt contributions with the \jpsi mesons originating from the decay vertices of beauty hadrons, which are typically separated from the PV, need to be subtracted in this analysis.
The prompt and nonprompt \jpsi mesons can be distinguished by exploiting the pseudoproper time~\cite{LHCb-PAPER-2011-003}
\begin{equation}
\label{eq:tz}
    t_z=\frac{z_{\jpsi}-z_{\text{PV}}}{p_z} \times m_{\jpsi},
\end{equation}
where $z_{\jpsi}$ and $z_{\text{PV}}$ are the positions of the \jpsi meson decay vertex and its associated PV along the beam axis $z$, 
$p_z$ the component of the \jpsi momentum along the $z$-axis, 
and $m_{\jpsi}$ the world-average value of the \jpsi mass~\cite{Workman:2022ynf}.
The uncertainty $\sigma_{t_z}$ is calculated by combining the uncertainties on $z_{\jpsi}$ and $z_{\text{PV}}$ since the uncertainty on $p_z$ is negligible in comparison.
Candidates with both \jpsi mesons having $-2<t_z<10\ps$ and $\sigma_{t_z}<0.3\ps$ are retained.
Finally, the four muon tracks are required to originate from the same PV, which reduces to a negligible level the number of candidates with the two \jpsi mesons originating from different $pp$ interactions.

\section{Cross-section determination}
\label{sec:Cross-section}
The di-\jpsi production cross-section is calculated as
\begin{equation}
\label{eq:csDiPsi}
    \sigma_{\text{di-}\jpsi}=
    \frac{N^{\text{corr}}}{\lum\times\BR^2(\jpsi\to\mumu)},
\end{equation}
where $N^{\text{corr}}$ is the signal yield after detection efficiency corrections,
${\lum=4.18\pm0.08\invfb}$ is the integrated luminosity measured using the van der Meer scan method~\cite{LHCb:2014vhh}, 
and ${\BR(\jpsi\to\mumu)=(5.961\pm0.033)\%}$~\cite{Workman:2022ynf} is the branching fraction of the $\jpsi\to\mumu$ decay.

\begin{figure}[tb]
  \centering
    \includegraphics[width=0.49\linewidth]{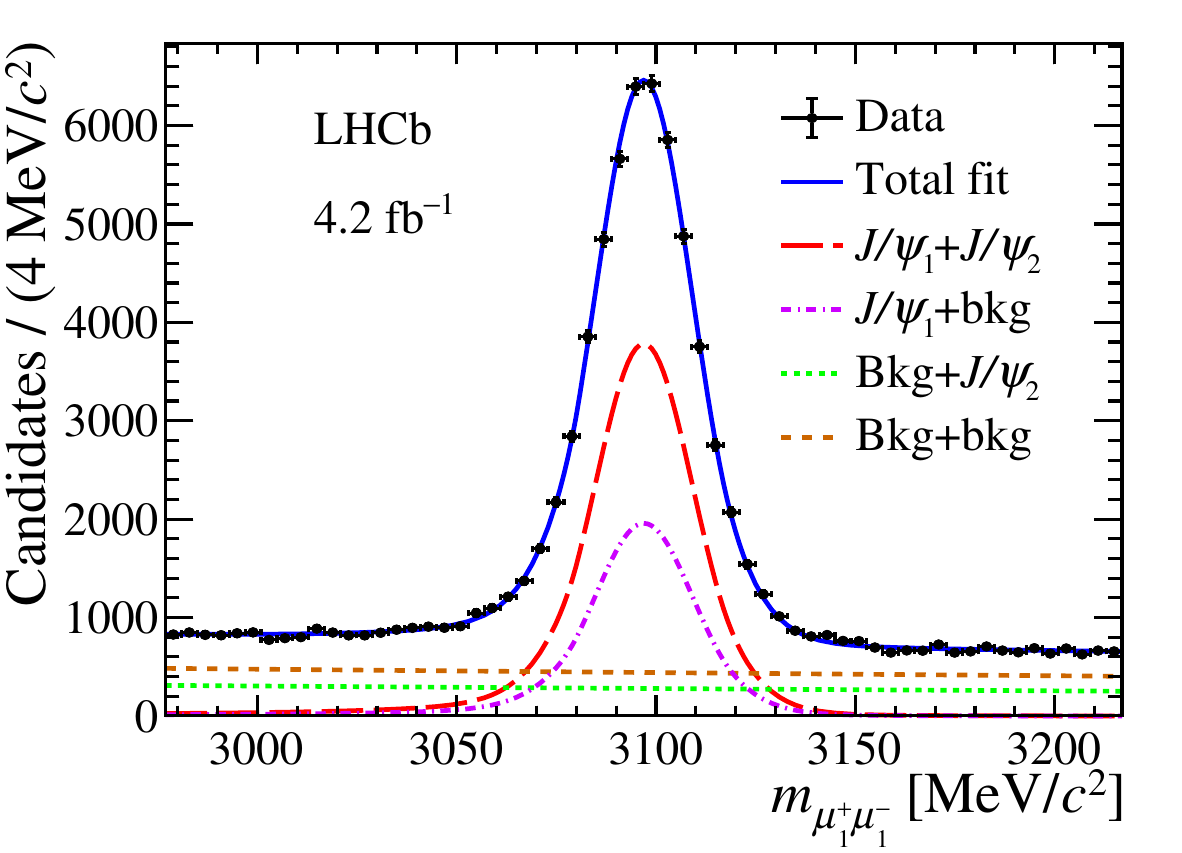}\put(-183,133){(a)}
    \hfil
    \includegraphics[width=0.49\linewidth]{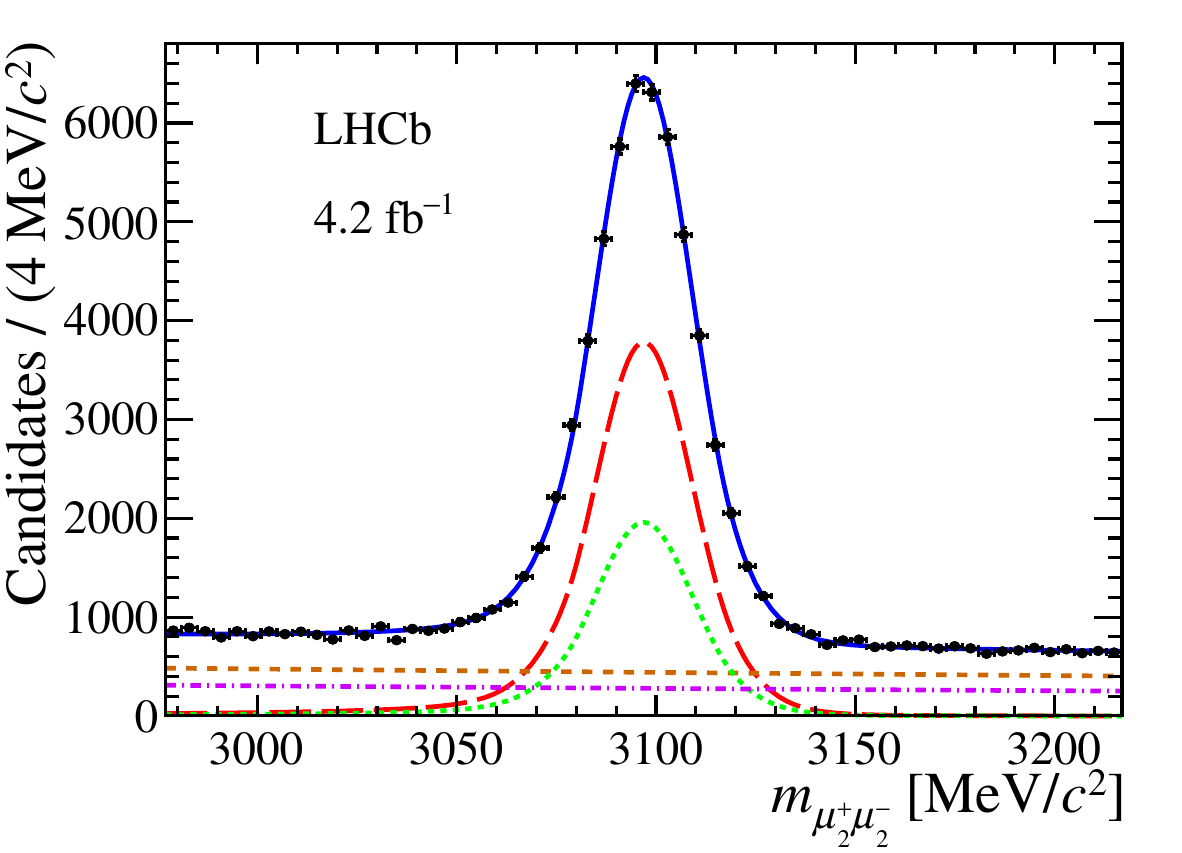}\put(-183,133){(b)}
  \caption{Invariant mass distributions of two $\mumu$ pairs (a) $m_{\mu_1^+\mu_1^-}$ and (b) $m_{\mu_2^+\mu_2^-}$ for di-\jpsi candidates together with the projections of the two-dimensional fit.}
  \label{fig:fitDoubleJpsi}
\end{figure}

The di-\jpsi signals are extracted by performing a two-dimensional~(2D) unbinned extended maximum likelihood fit to the distribution of the two \jpsi meson masses, $(m_{\mu^+_1\mu^-_1},m_{\mu^+_2\mu^-_2})$.
The two $\jpsi\to\mumu$ decays are labelled as $\jpsi_1\to\mu^+_1\mu^-_1$ and $\jpsi_2\to\mu^+_2\mu^-_2$ at random.
There are four components of the 2D mass distribution: a signal di-\jpsi decay, a true $\jpsi_1\to\mu^+_1\mu^-_1$ decay with a dimuon background $\mu^+_2\mu^-_2$, a dimuon background $\mu^+_1\mu^-_1$ with a true $\jpsi_2\to\mu^+_2\mu^-_2$ decay, and the association of two combinatorial dimuon backgrounds.
The second and the third components have the equal fraction because the mass distributions of the two \jpsi mesons are symmetric.
For the mass distribution of each \jpsi meson, 
the signal component is described by the sum of a double-sided Crystal Ball (DSCB) function~\cite{Skwarnicki:1986xj} with asymmetric tails and a Gaussian function with a common mean value but different widths.
The tail parameters of the DSCB function, the fraction of the DSCB function and the ratio between the two widths are fixed from simulation.
Only the common mean value and the width of the DSCB function are left as free parameters.
The distribution of the combinatorial dimuon background component is modelled with an exponential function.
Figure~\ref{fig:fitDoubleJpsi} shows the mass distributions for the two $\mumu$ pairs and the projections of the fit result on $m_{\mu^+_1\mu^-_1}$ and $m_{\mu^+_2\mu^-_2}$.

\begin{figure}[tb]
  \centering
    \includegraphics[width=0.49\linewidth]{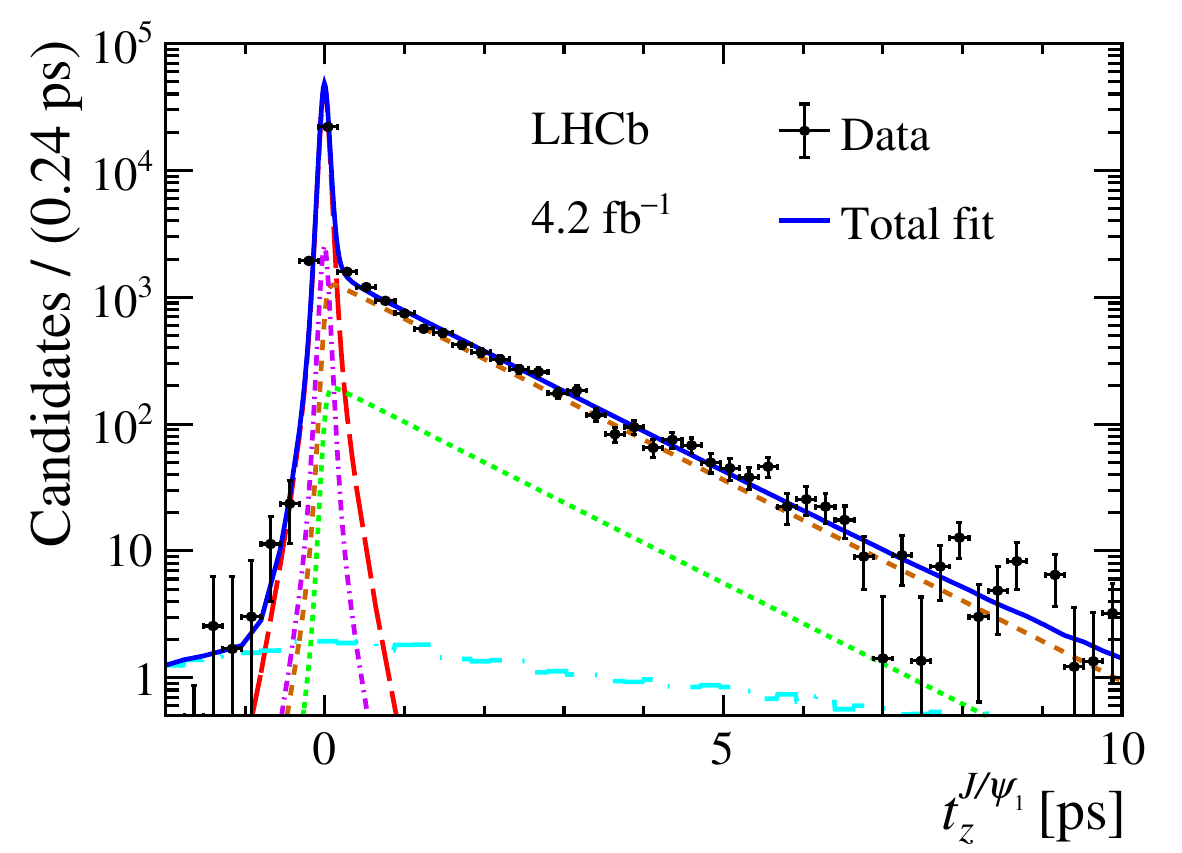}\put(-183,133){(a)}
    \hfil
    \includegraphics[width=0.49\linewidth]{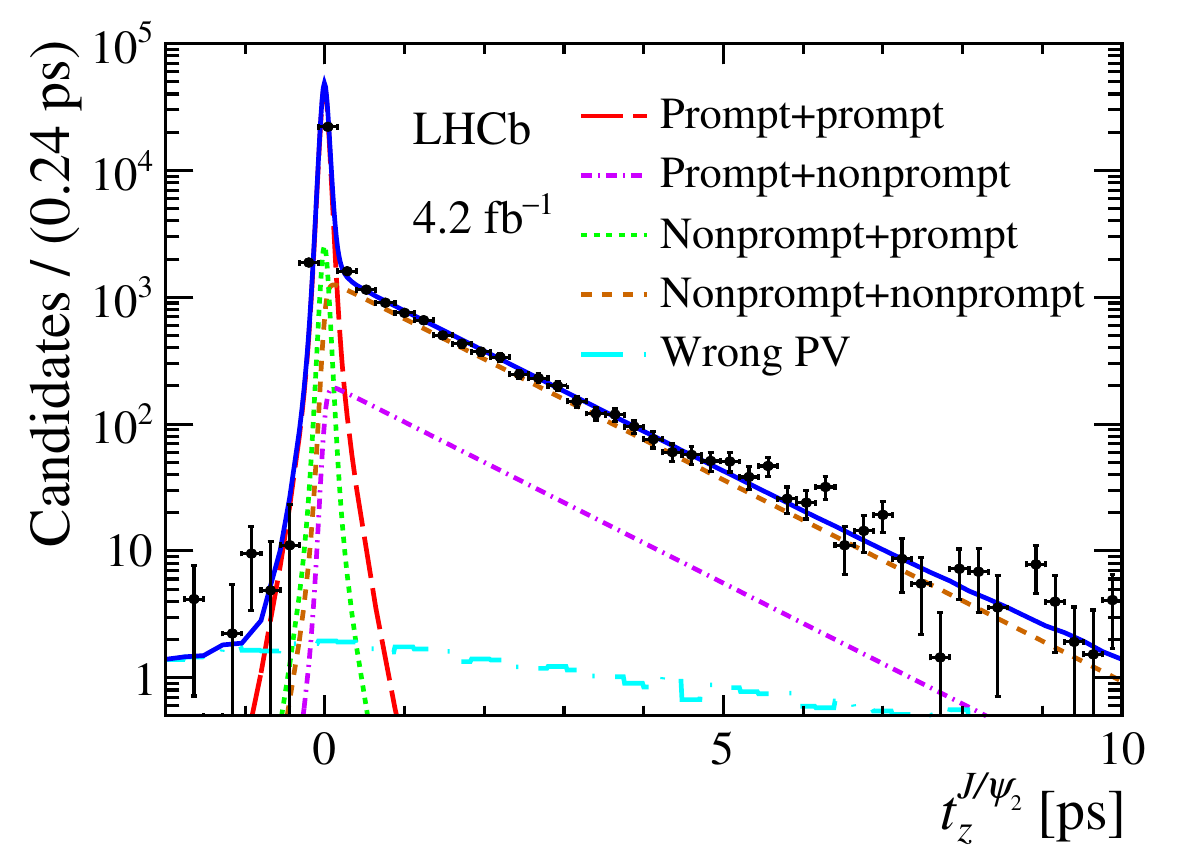}\put(-183,133){(b)}
  \caption{Distributions of (a) $t_z^{\jpsi_1}$ and (b) $t_z^{\jpsi_2}$ for di-\jpsi signals with backgrounds subtracted. The projections of the fit result are overlaid.}
  \label{fig:fitDoubleJpsiTz}
\end{figure}

The subtraction of the nonprompt contributions is performed using a 2D unbinned extended maximum likelihood fit to the $t_z$ distribution of each \jpsi meson of the di-\jpsi signals with backgrounds subtracted using the \sPlot method~\cite{Pivk:2004ty}, taking $m_{\mu_1^+\mu_1^-}$ and $m_{\mu_2^+\mu_2^-}$ as the discriminating variables.
The true $t_z$ distribution of prompt \jpsi mesons is expected to follow a Dirac delta function $\delta(t_z)$, 
while that of nonprompt \jpsi mesons should follow an exponential function.
These are convolved with the sum of two Gaussian functions to model the detector resolution.
The parameters of the resolution function are free parameters in the fit.
A third component describes candidates with incorrectly associated PVs and is modelled using a binned histogram extracted from data by calculating $t_z$ with one \jpsi meson associated to the closest PV in the next event.
Figure~\ref{fig:fitDoubleJpsiTz} shows the $t_z$ distributions for the two \jpsi mesons and the projections of the 2D fit result on $t_z^{\jpsi_1}$ and $t_z^{\jpsi_2}$.
The yield of prompt di-\jpsi signal is $(2.187\pm0.020)\times10^4$, accounting for around 68\% of the total di-\jpsi signal yield.

Since the kinematic distribution of the di-\jpsi signal is unknown a priori, the efficiency correction is performed on a per-event basis as
\begin{equation}
\label{eq:corrN}
    N^{\text{corr}}=\sum_i \frac{\omega_i}{\etot_i},
\end{equation}
where $i$ is the event index, 
$\omega_i$ is the \sPlot weight corresponding to the component of prompt di-\jpsi signal, 
and $\etot_i$ is the total efficiency for each candidate.
Since no information related to the correlation of the two \jpsi mesons is used during the reconstruction and selection,
the detection efficiency of the di-\jpsi candidate can be factorised into that of the two \jpsi mesons,
which are determined from simulation as functions of the $\pt$ and $y$ of the \jpsi mesons.
The efficiency $\etot_i$ is the product of the geometrical acceptance $\eacc_i$, the reconstruction and selection efficiency $\erecsel_i$, the particle identification~(PID) efficiency $\epid_i$, and the trigger efficiency $\etri_i$, giving
\begin{equation}
    \etot_i = \eacc_i\times\erecsel_i \times\epid_i\times\etri_i.
\end{equation}
The efficiencies $\eacc_i$, $\erecsel_i$ and $\epid_i$ of a di-\jpsi candidate factorise as the product of that of the two \jpsi mesons, \ie
\begin{equation}
    \varepsilon_i(\text{di-}\jpsi)=\varepsilon_i(\jpsi_1) \times\varepsilon_i(\jpsi_2),
\end{equation}
while the trigger efficiency $\etri_i$ of each di-\jpsi candidate factorises as
\begin{equation}
    \etri_i(\text{di-}\jpsi)= 1-\left(1-\varepsilon^{\text{L0\&HLT1}}_i(\jpsi_1)\right)\left(1-\varepsilon^{\text{L0\&HLT1}}_i(\jpsi_2)\right),
\end{equation}
where $\varepsilon^{\text{L0\&HLT1}}_i$ is the L0 and HLT1 trigger efficiency.
The offline selection criterion is tighter than the HLT2 requirements, making the HLT2 trigger fully efficient with respect to offline selected candidates.
The track reconstruction efficiency, which is part of $\erecsel_i$, and the PID efficiency $\epid_i$ are calibrated using data-driven techniques to avoid known discrepancies between the simulation and data~\cite{LHCb-DP-2013-002,LHCb-DP-2013-001}.
With the detection efficiency corrected, the signal yield is determined to be $N^{\text{corr}}=(2.43\pm0.04)\times 10^5$, where the statistical uncertainty is verified by a bootstrapping approach~\cite{Efron:1979bxm}.

\section{Systematic uncertainties}
Systematic uncertainties are studied and summarised in Table~\ref{table:sysDiPsi}.
The uncertainty due to imperfect modelling of the \jpsi mass distribution is estimated by using an alternative function for the \jpsi signal component in the fit.
A model derived from the simulation using kernel density estimation~\cite{Cranmer:2000du} is used instead.
The relative variation of the extracted signal yield is 1.7\%, which is taken as the systematic uncertainty.
The uncertainty on the determination of the nonprompt contribution is evaluated by using an alternative variable for the discrimination.
The $\log(\chisqip)$ of the \jpsi mesons is used instead of $t_z$, 
where \chisqip is defined as the difference in the vertex-fit \chisq of the associated PV reconstructed with and without the \jpsi meson under consideration.
The relative deviation from the nominal result, 2.4\%, is taken as the systematic uncertainty.
For the measurement of differential cross-sections, these two uncertainties are taken as common to all kinematic intervals.

For a very small fraction of the di-\jpsi candidates, the \jpsi mesons may be associated to a wrong PV in two cases.
In the first, the true PV is reconstructed but one of the two \jpsi candidates is associated to a wrong PV, 
while in the second the true PV is not reconstructed and both \jpsi mesons are associated to the same reconstructed PV in this event.
For the first case, two \jpsi mesons are associated to different PVs and thus are rejected by the selection.
The fraction of the first case is estimated using a simulated sample of ${\OneS\to\jpsi\jpsi\gamma}$ decays to be $(0.65\pm0.02)\%$.
The second case is the wrong-PV component in the 2D $t_z$ fit, and its fraction is $(0.16\pm0.08)\%$.
The total fraction of wrong-PV candidates adds up to 0.8\%, which is taken as the systematic uncertainty.
For the differential cross-sections, the fraction is studied in several intervals of the di-\jpsi rapidities using the same method.
The maximum fraction, 1.5\%, is conservatively taken as the uncertainty common to all kinematic intervals.

The uncertainty due to the finite sample size of the calibration samples used for efficiency determination is propagated to the final result using pseudoexperiments.
It is determined to be 0.2\% and varies up to 1.1\% depending on the kinematic intervals for the differential cross-sections.
The binning scheme that is used to determine the efficiencies could bias the signal yield $N^{\text{corr}}$.
For the PID efficiency, this effect is estimated by varying the binning schemes of the PID calibration sample.
For the remaining efficiency terms, an alternative kernel density estimation approach~\cite{Cranmer:2000du} is used to determine the efficiencies as functions of $(\pt^{\jpsi},y_{\jpsi})$.
The relative difference in the cross-sections between the default and alternative approaches is 1.5\% and quoted as the systematic uncertainty.
For the measurement of differential cross-sections, it varies up to 6.6\% depending on the kinematic intervals.
The uncertainty on the track reconstruction efficiency consists of the statistical uncertainty due to the limited size of the calibration samples,
which is estimated from pseudoexperiments to be 1.2\% for di-\jpsi candidates,
and the uncertainty due to the dependence of calibration factors on the event multiplicity,
which is 0.8\% per track.
The two terms are added in quadrature to 3.4\%.
For the differential cross-sections, the first term varies up to 3.8\%, while the second is considered to be common to all kinematic intervals.
The trigger efficiency determined from the simulation is validated with the prompt \jpsi data, using a subset of events that fulfil the trigger requirement with the \jpsi signals excluded~\cite{LHCb-DP-2012-004}.
The relative difference in the di-\jpsi cross-section calculated using the trigger efficiencies from the data and the simulation is 0.7\%, and is taken as the systematic uncertainty.
For the differential cross-sections, the uncertainty on the trigger efficiency varies up to 7.9\% depending on the kinematic intervals.

The uncertainty on the ${\jpsi\to\mumu}$ branching fraction leads to an uncertainty of 1.1\% on the di-\jpsi production cross-section.
The relative uncertainty on the luminosity is determined to be 2.0\%.
With these uncertainties due to independent effects added in quadrature, the total systematic uncertainty on the di-\jpsi production cross-section is determined to be 5.4\%.

\begin{table}[tb]
    \centering
    \caption{Summary of the systematic uncertainties on the measurement of the di-\jpsi production cross-section. The total systematic uncertainty is a quadratic sum of these uncertainties.}
    \label{table:sysDiPsi}
    \begin{tabular}{lc}
	\hline
	Source                       		& Uncertainty (\%) \\
	\hline
	Signal mass model               	& 1.7   \\
	Nonprompt contribution              & 2.4   \\
	Wrong PV association         		& 0.8   \\
	Calibration sample statistics       & 0.2   \\
	Efficiency determination            & 1.5   \\
	Track reconstruction efficiency     & 3.4   \\
	Trigger efficiency                  & 0.7   \\
	Branching fraction           		& 1.1   \\
	Luminosity                   		& 2.0   \\ 
	\hline
	Total                               & 5.4   \\
	\hline
    \end{tabular}
\end{table}

\section{Production cross-sections}
\label{sec:Production}
The cross-section of the di-\jpsi production with both \jpsi mesons in the fiducial range ${\pt<14\gevc}$ and ${2.0<y<4.5}$ is measured to be
\begin{equation*}
    \sigma_{\text{di-}\jpsi}= 16.36\pm0.28~(\text{stat})\pm0.88~(\text{syst})\nb, \\
\end{equation*}
where the first uncertainty is statistical and the second systematic, assuming negligible polarisation of the \jpsi mesons.
The detection efficiency for the \jpsi mesons depends on their polarisation. In the helicity frame, there is a strong dependence on the polarisation parameter $\lambda_{\theta}$~\cite{LHCb-PAPER-2013-008,LHCB-PAPER-2015-037}.
For instance, when $\lambda_{\theta}$ is assumed to be $+0.2$ ($-0.2$) for both \jpsi mesons, the di-\jpsi production cross-section changes by $+6.2\%$ ($-6.3\%$) evaluated from simulation.

\begin{figure}[tb]
  \centering
    \includegraphics[width=0.41\linewidth]{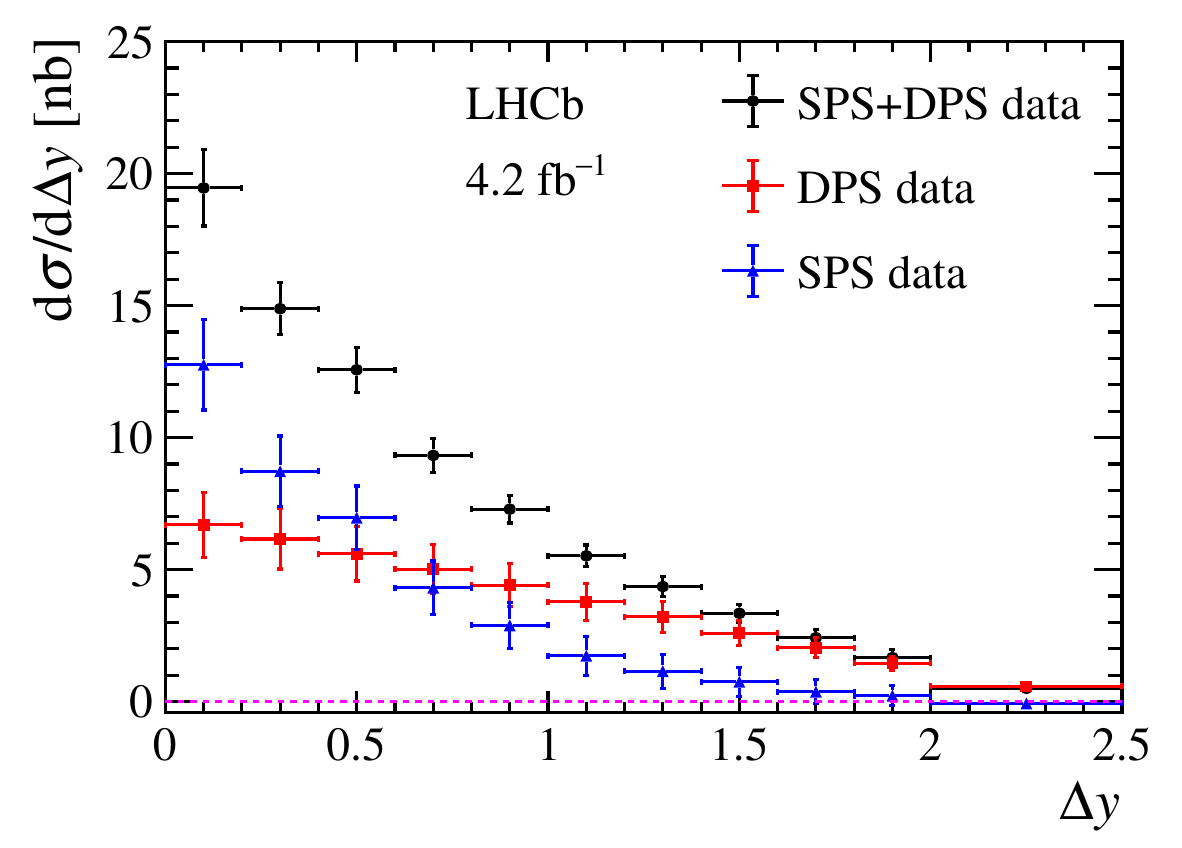}\put(-150,110){(a)}
    \includegraphics[width=0.41\linewidth]{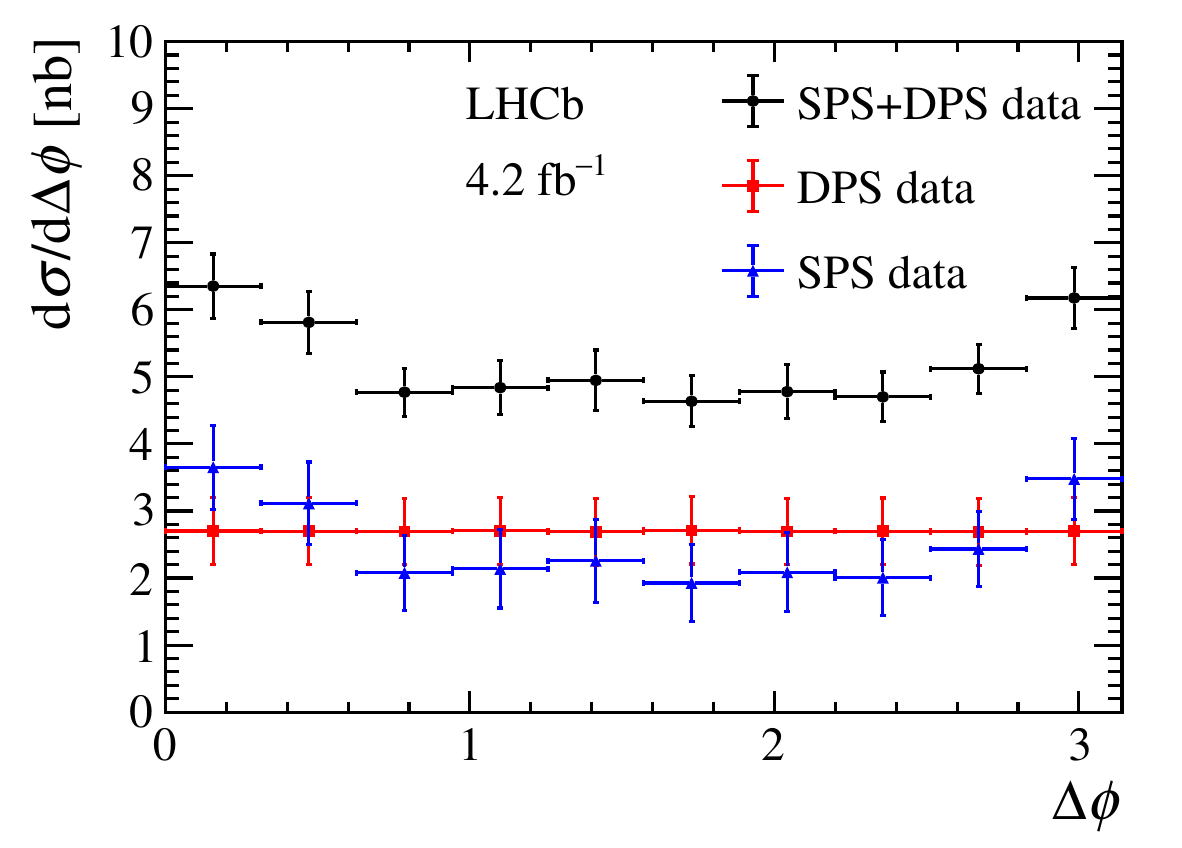}\put(-150,110){(b)}
    \\
    \includegraphics[width=0.41\linewidth]{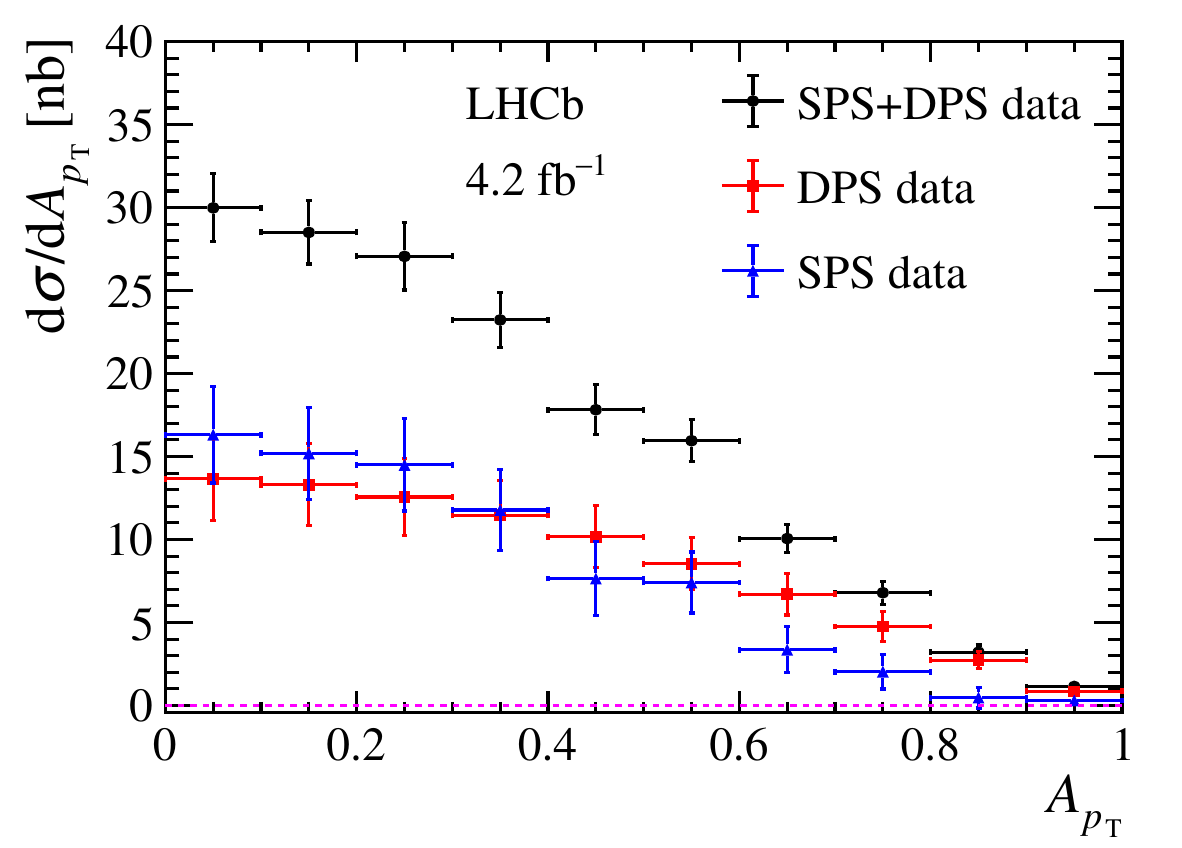}\put(-150,110){(c)}
    \includegraphics[width=0.41\linewidth]{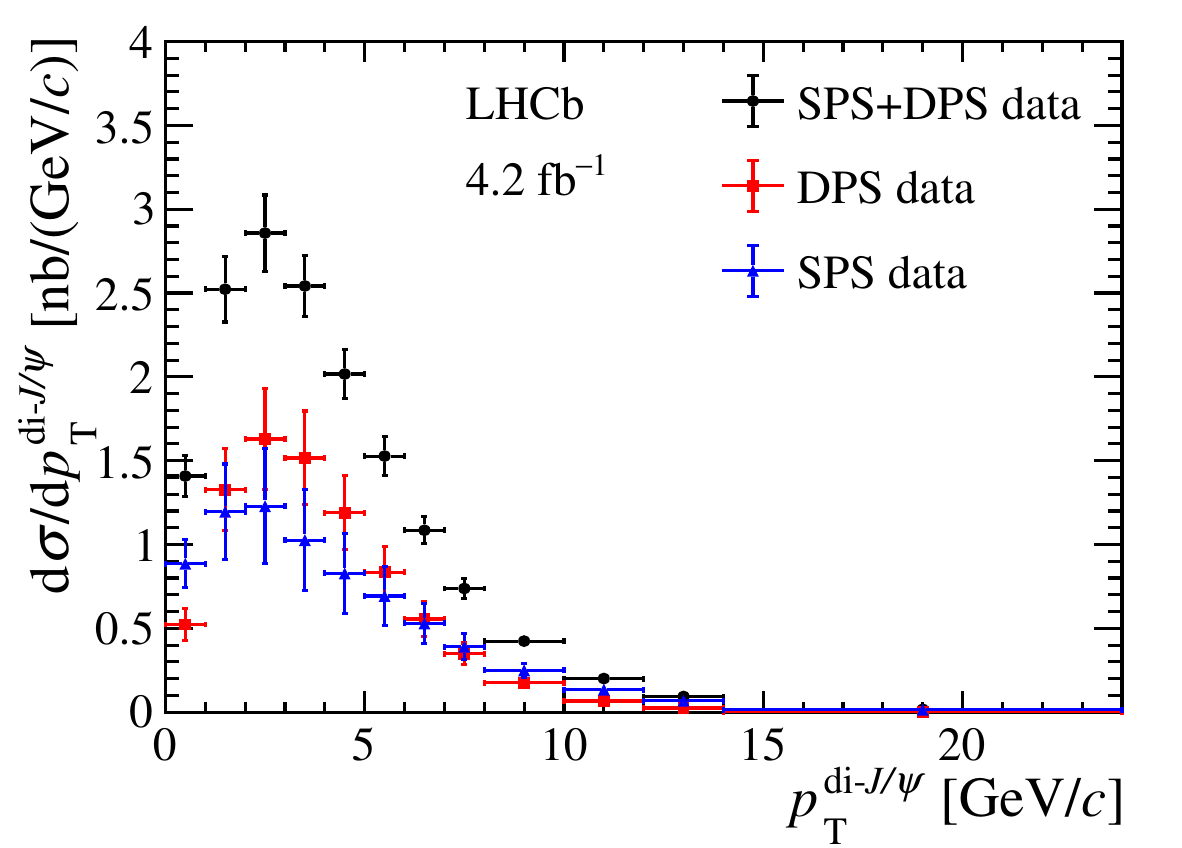}\put(-150,110){(d)}
    \\
    \includegraphics[width=0.41\linewidth]{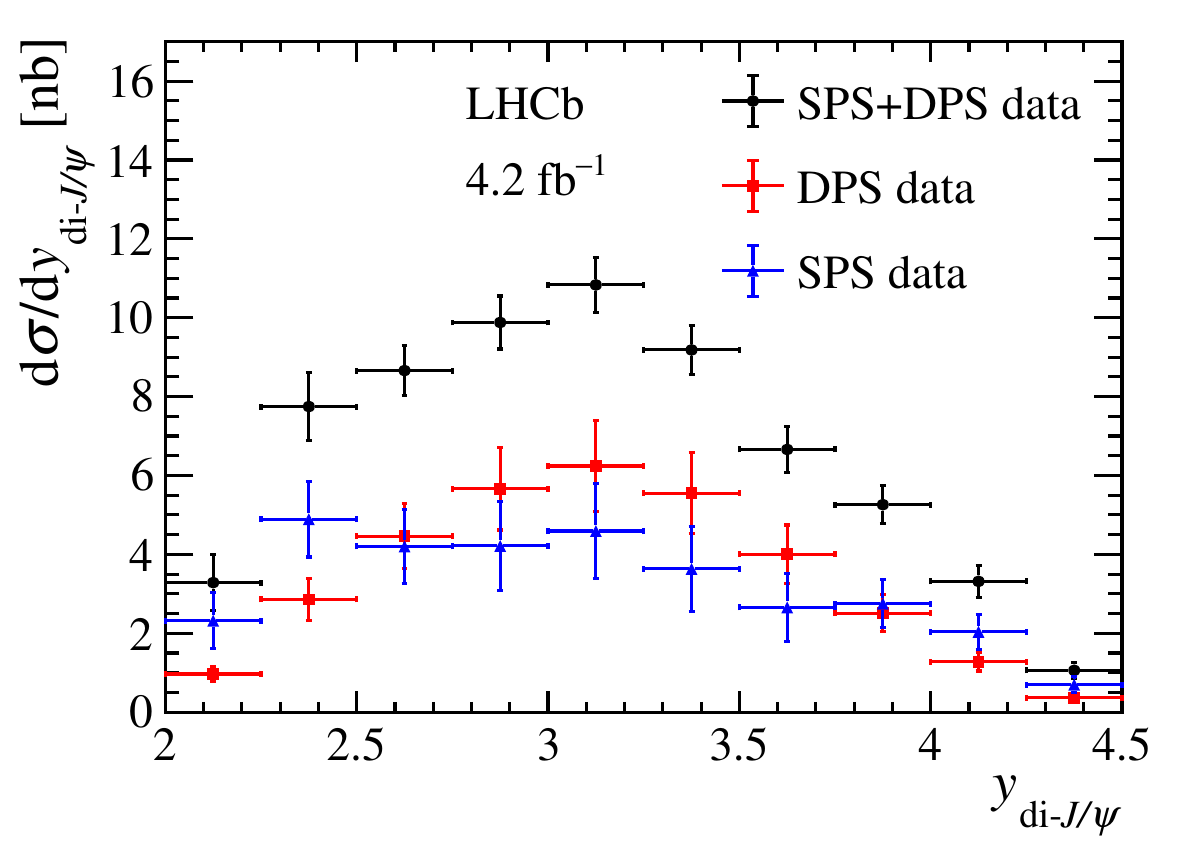}\put(-150,110){(e)}
    \includegraphics[width=0.41\linewidth]{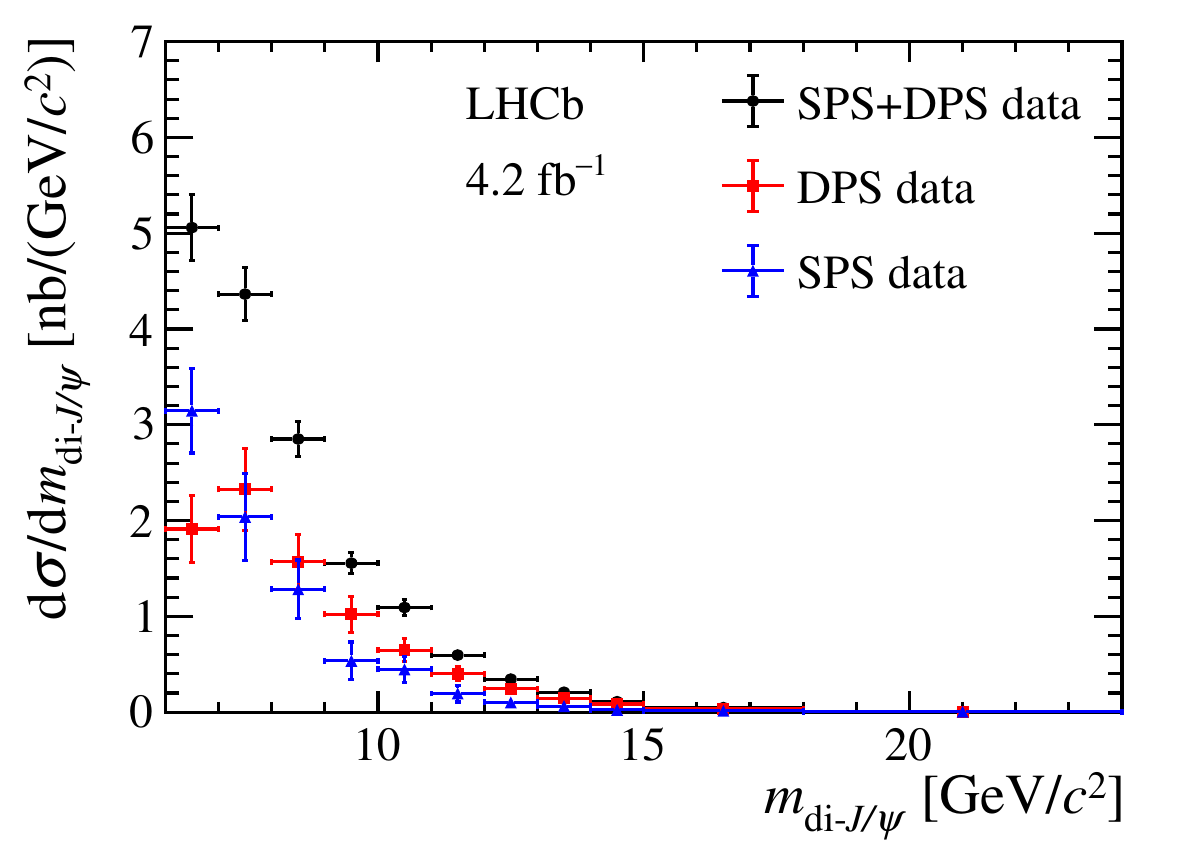}\put(-150,110){(f)}
    \\
    \includegraphics[width=0.41\linewidth]{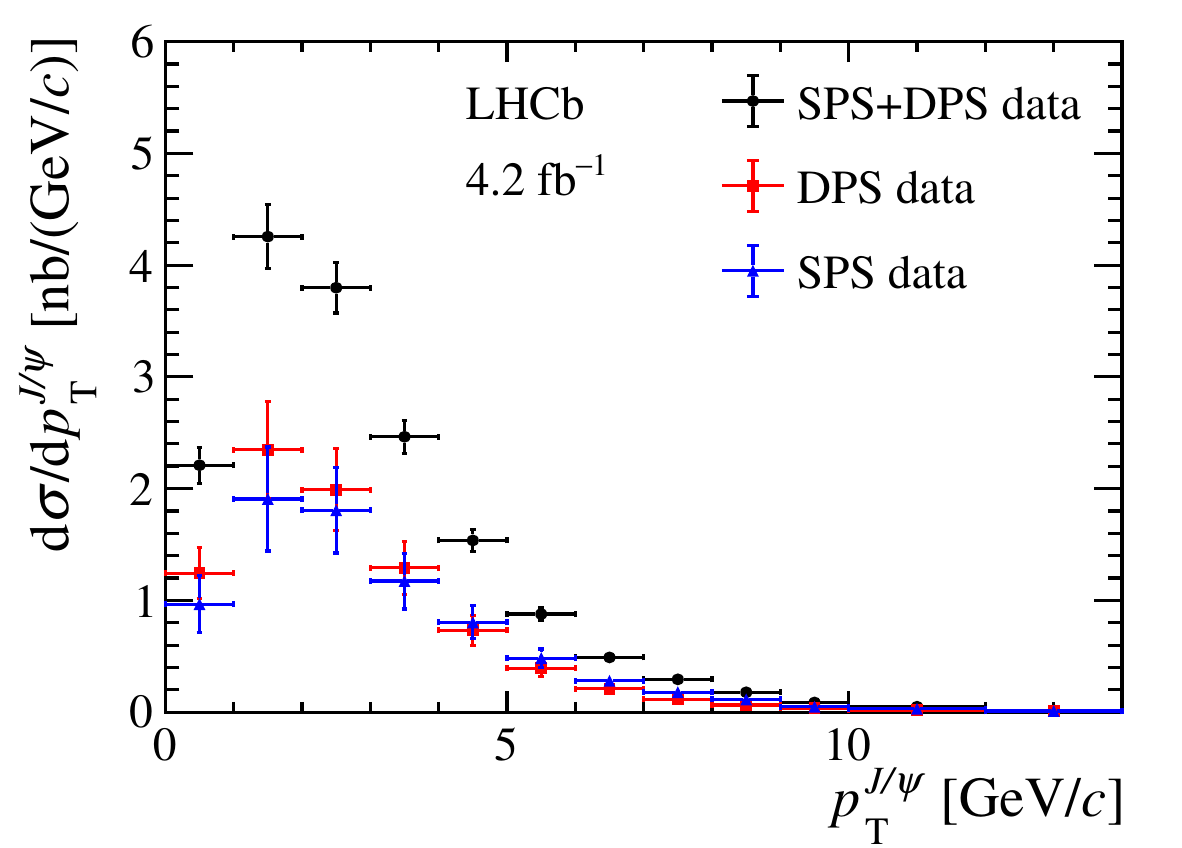}\put(-150,110){(g)}
    \includegraphics[width=0.41\linewidth]{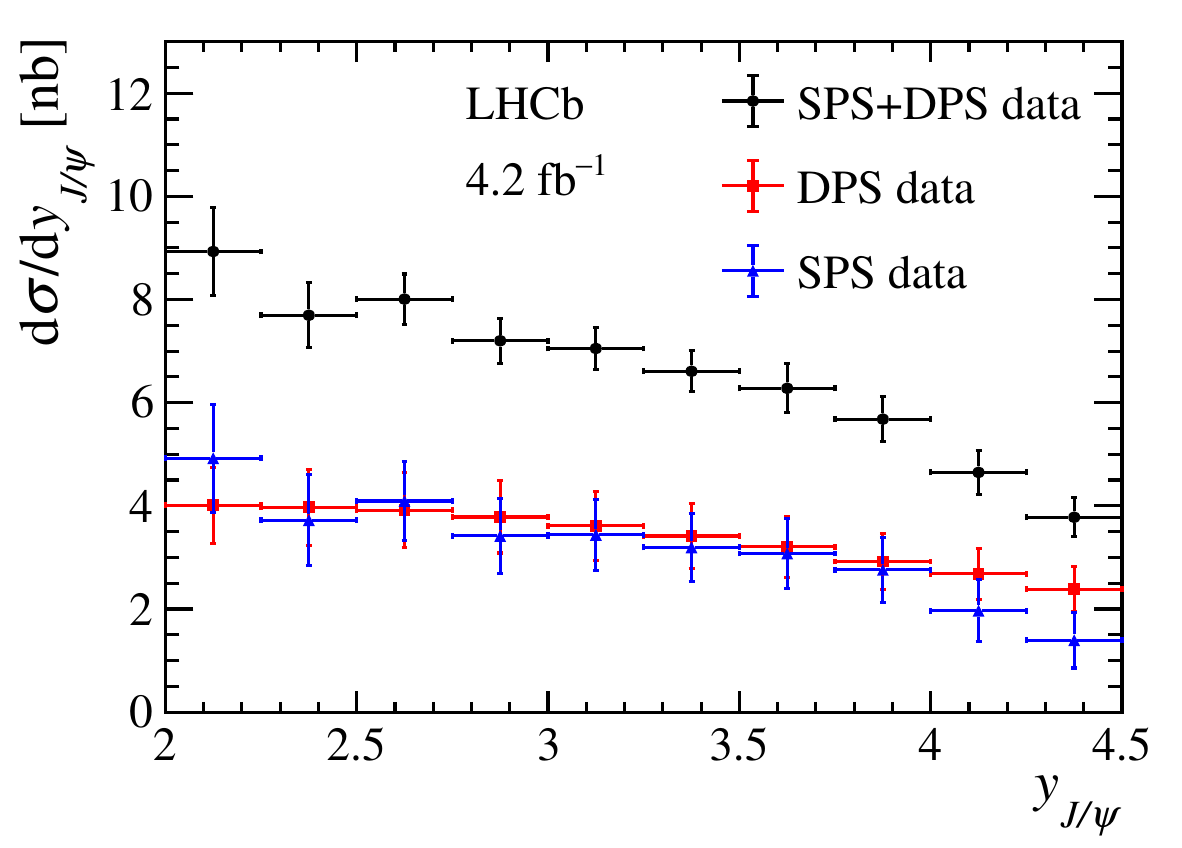}\put(-150,110){(h)}
  \caption{Differential cross-section of di-\jpsi production for SPS+DPS, DPS and SPS as a function of (a) $\Delta y$, (b) $\Delta\phi$, (c) $\mathcal{A}_{\pt}$, (d) $\pt^{\text{di-}\jpsi}$, (e) $y_{\text{di-}\jpsi}$, (f) $m_{\text{di-}\jpsi}$, (g) $\pt^{\jpsi}$ and (h) $y_{\jpsi}$. The error bars represent the statistical and systematic uncertainties added in quadrature. The purple dashed lines in (a) and (c) indicate the baseline of zero.}
  \label{fig:csDiffSPS}
\end{figure}

The differential di-\jpsi production cross-section as a function of a kinematic variable $u$ is measured as
\begin{equation}
\label{eq:csDiffDiPsi}
    \frac{\deriv\sigma_{\text{di-}\jpsi}}{\deriv u}=
    \frac{\Delta N^{\text{corr}}(\text{di-}\jpsi)}{\lum\times\BR^2(\jpsi\to\mumu)\times\Delta u},
\end{equation}
where $\Delta N^{\text{corr}}$ is the efficiency-corrected signal yield in an interval of the variable $u$, and $\Delta u$ is the interval width.
In this analysis the cross-sections are reported as functions of:
the absolute difference in rapidity between the two \jpsi mesons $\Delta y$;
the absolute difference in the azimuthal angle $\phi$, defined in the laboratory frame, between the two \jpsi mesons $\Delta\phi$;
the transverse momentum asymmetry $\mathcal{A}_{\pt}$ of the two \jpsi mesons, defined as
\begin{equation}
    \mathcal{A}_{\pt}=\left|\frac{\pt^{\jpsi_1}-\pt^{\jpsi_2}}{\pt^{\jpsi_1}+\pt^{\jpsi_2}}\right|;
\end{equation}
the transverse momentum $\pt^{\text{di-}\jpsi}$, rapidity $y_{\text{di-}\jpsi}$ and invariant mass $m_{\text{di-}\jpsi}$ of the di-\jpsi signals;
and the transverse momentum $\pt^{\jpsi}$ and rapidity $y_{\jpsi}$ of either \jpsi meson.
The binning scheme is chosen to have adequate and approximately even signal yield in each category.
The differential cross-section as a function of $\pt^{\jpsi}$ ($y_{\jpsi}$) is taken as the average of the two distributions of $\pt^{\jpsi_1}$ ($y_{\jpsi_1}$) and $\pt^{\jpsi_2}$ ($y_{\jpsi_2}$), taking advantage of the symmetry between the two \jpsi mesons.
The 2D mass fit and the 2D pseudoproper time fit are performed independently for each kinematic interval to subtract the combinatorial backgrounds and the nonprompt contributions, respectively.
The measured differential cross-sections of the di-\jpsi production
are shown in Figure~\ref{fig:csDiffSPS} as black data points (SPS+DPS),
and summarised in Tables~\ref{table:csDoubleJpsiDY}--\ref{table:csDoubleJpsiJpsiY} in 
Appendix~\ref{sec:Tables}.

\section{Separation of DPS and SPS contributions}
\label{sec:Separation}
\begin{figure}[tb]
  \centering
  \includegraphics[width=0.6\linewidth]{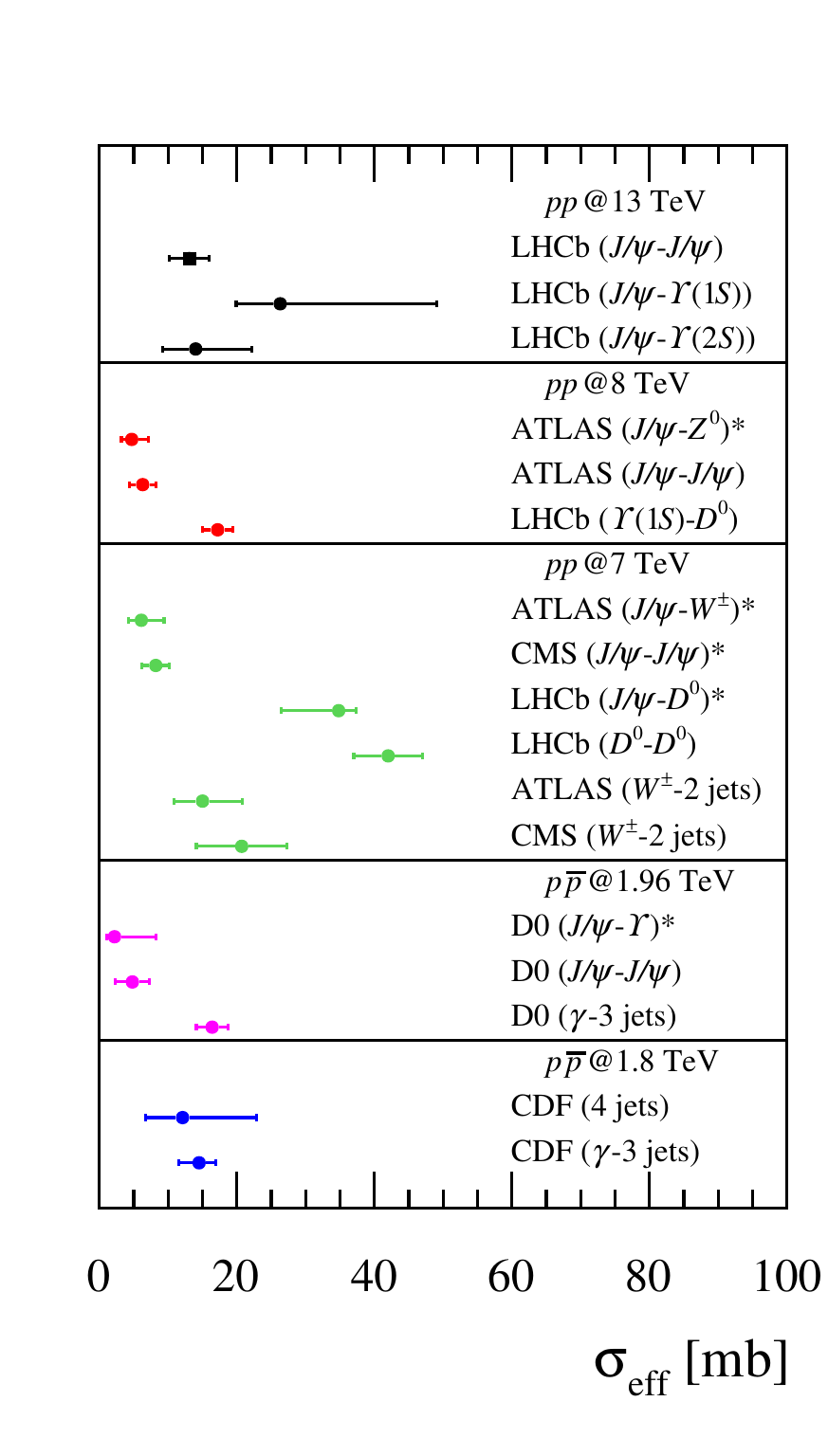}
  \caption{Summary of $\sigma_{\text{eff}}$ measurements in $pp$ or $\proton\antiproton$ collisions~\cite{LHCb:2023qgu,ATLAS:2014ofp,ATLAS:2016ydt,LHCb:2015wvu,ATLAS:2014yjd,CMS:2014cmt,LHCb:2012aiv,ATLAS:2013aph,CMS:2013huw,D0:2015dyx,D0:2014vql,D0:2009apj,CDF:1993sbj,CDF:1997yfa}. The legend entries marked with an asterisk are taken from a third-party calculation based on the original experimental result~\cite{Lansberg:2016rcx,Lansberg:2017chq,Lansberg:2014swa,Shao:2020kgj,Shao:2016wor}.}
  \label{fig:sigmaEff}
\end{figure}

The distributions of the rapidity difference ($\Delta y$) between the two \jpsi mesons have different shapes for the SPS and DPS processes~\cite{Kom:2011bd,D0:2014vql,ATLAS:2016ydt},
so the DPS contribution can be extracted using the $\Delta y$ distribution with a data-driven template for the DPS process.
The shape of the DPS component is obtained by combining two uncorrelated \jpsi mesons whose distributions follow the measured differential production cross-section of the single prompt \jpsi~\cite{LHCB-PAPER-2015-037},
assuming they are both uniformly distributed over the azimuthal angle $\phi$.
According to the NRQCD predictions~\cite{Lansberg:2019fgm, Sun:2014gca, Likhoded:2016zmk, PhysRevD.84.054012},
the SPS contribution to the di-\jpsi production in the range $1.8<\Delta y<2.5$ is negligible.
The normalisation of the DPS contribution is thus determined in this range, while the remaining contribution is assigned to SPS.
For the extraction of the DPS contribution, three sources of systematic uncertainties are considered:
the uncertainty due to possible SPS remnant in the $1.8<\Delta y<2.5$ range, which is studied by varying the $\Delta y$ range for normalisation and estimated to be 3.5\%;
the uncertainty on the DPS template due to the binning scheme of the prompt \jpsi differential cross-sections, which is estimated to be 3.3\%, determined through the relative difference between the results with and without interpolation across the intervals;
the uncertainty propagated from the prompt \jpsi production cross-section, which is 1.8\%.
Consequently, the DPS cross-section of the di-\jpsi production with both \jpsi mesons in the fiducial range ${\pt<14\gevc}$ and ${2.0<y<4.5}$ is determined to be
\begin{equation*}
    \sigma^{\text{DPS}}_{\text{di-}\jpsi}= 8.6\pm1.2~(\text{stat})\pm1.0~(\text{syst})\nb.
\end{equation*}
According to Eq.~\ref{eq:pocket}, the effective cross-section $\sigma_{\text{eff}}$ is measured to be
\begin{equation*}
    \sigma_{\text{eff}}=\frac12\frac{\sigma^2_{\jpsi}}{\sigma^{\text{DPS}}_{\text{di-}\jpsi}}=13.1\pm1.8~(\text{stat})\pm2.3~(\text{syst})\mbarn,
\end{equation*}
where the prompt \jpsi production cross-section $\sigma_{\jpsi}$ in the range ${0<\pt<14\gevc}$ and ${2.0<y<4.5}$ is ${\sigma_{\jpsi}=15.03\pm0.03~(\text{stat})\pm0.94~(\text{syst})\mub}$~\cite{LHCB-PAPER-2015-037}, and the systematic uncertainties on $\sigma^{\text{DPS}}_{\text{di-}\jpsi}$ and $\sigma_{\jpsi}$ are treated as uncorrelated.
The $\sigma_{\text{eff}}$ result is compatible with existing measurements from different experiments in $pp$ and $\proton\antiproton$ collisions, as shown in Figure~\ref{fig:sigmaEff}.
With the DPS cross-section subtracted, the SPS cross-section of di-\jpsi production with both \jpsi mesons in the fiducial range ${\pt<14\gevc}$ and ${2.0<y<4.5}$ is determined to be
\begin{equation*}
    \sigma^{\text{SPS}}_{\text{di-}\jpsi}= 7.9\pm1.2~(\text{stat})\pm1.1~(\text{syst})\nb.
\end{equation*}

The differential di-\jpsi production cross-sections are shown in Figure~\ref{fig:csDiffSPS} with the DPS and SPS contributions separated.
The differential cross-sections for DPS are obtained by normalising the data-driven DPS template to the total DPS cross-section in the fiducial range, and the uncertainties are propagated from the total DPS cross-section.
The extracted differential SPS cross-sections are listed in Tables~\ref{table:csDoubleJpsiDYSPS}--\ref{table:csDoubleJpsiJpsiYSPS} in Appendix~\ref{sec:Tables}.
By definition the SPS and DPS components are anti-correlated.
In general, the difference in distributions between DPS and SPS is due to the fact that the kinematics of two \jpsi mesons are uncorrelated for DPS while correlated for SPS.
Figure~\ref{fig:csDiffSPS}(a) shows that the $\Delta y$ distribution for the DPS process is wider than that for the SPS process.
As shown in Figure~\ref{fig:csDiffSPS}(b), the $\Delta \phi$ distribution for SPS  peaks at $\Delta\phi=0$ and $\pi$, while the DPS distribution is flat because two uncorrelated \jpsi mesons are assumed to be uniformly distributed over the angle $\phi$ in the data-driven template.
The distributions of $\mathcal{A}_{\pt}$ and $m_{\text{di-}\jpsi}$ for DPS are both slightly wider than those for SPS, as shown in Figures~\ref{fig:csDiffSPS}(c) and~\ref{fig:csDiffSPS}(f).

\begin{figure}[tb]
  \centering
    \includegraphics[width=0.41\linewidth]{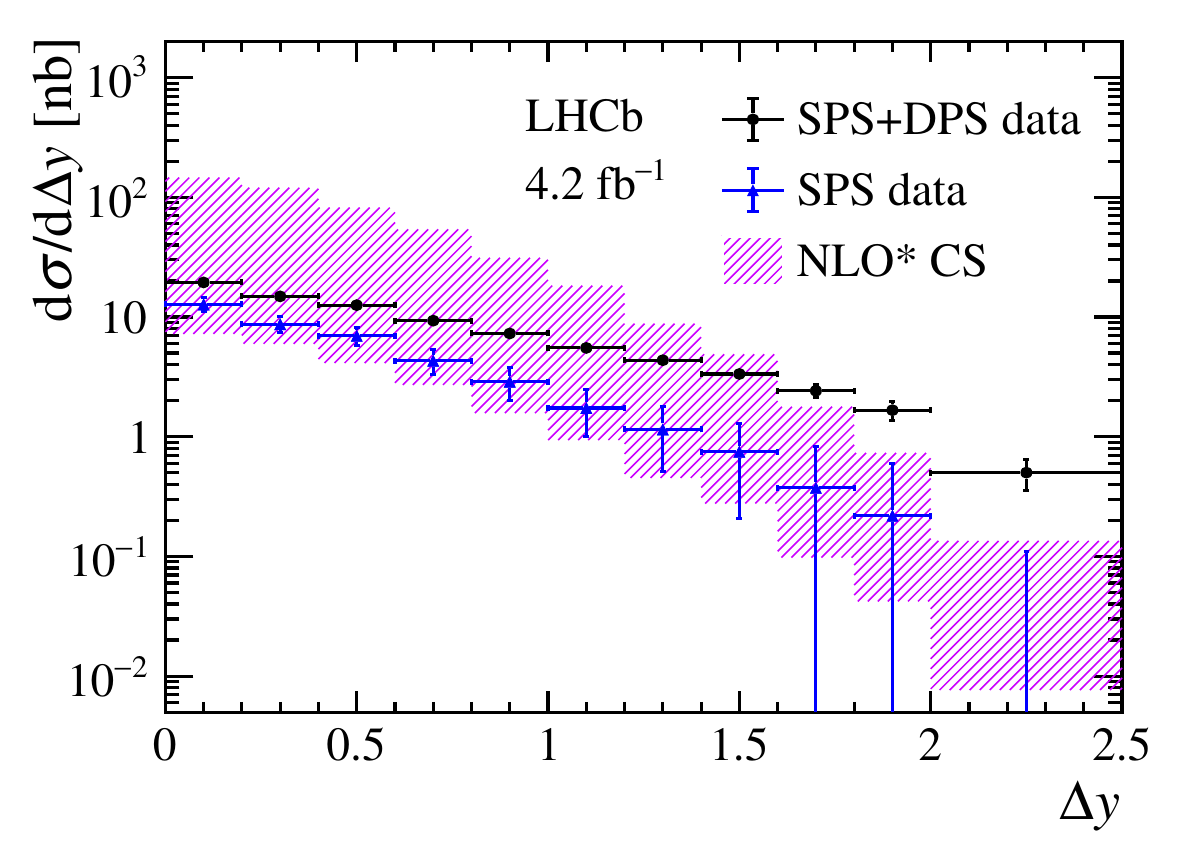}\put(-150,110){(a)}
    \includegraphics[width=0.41\linewidth]{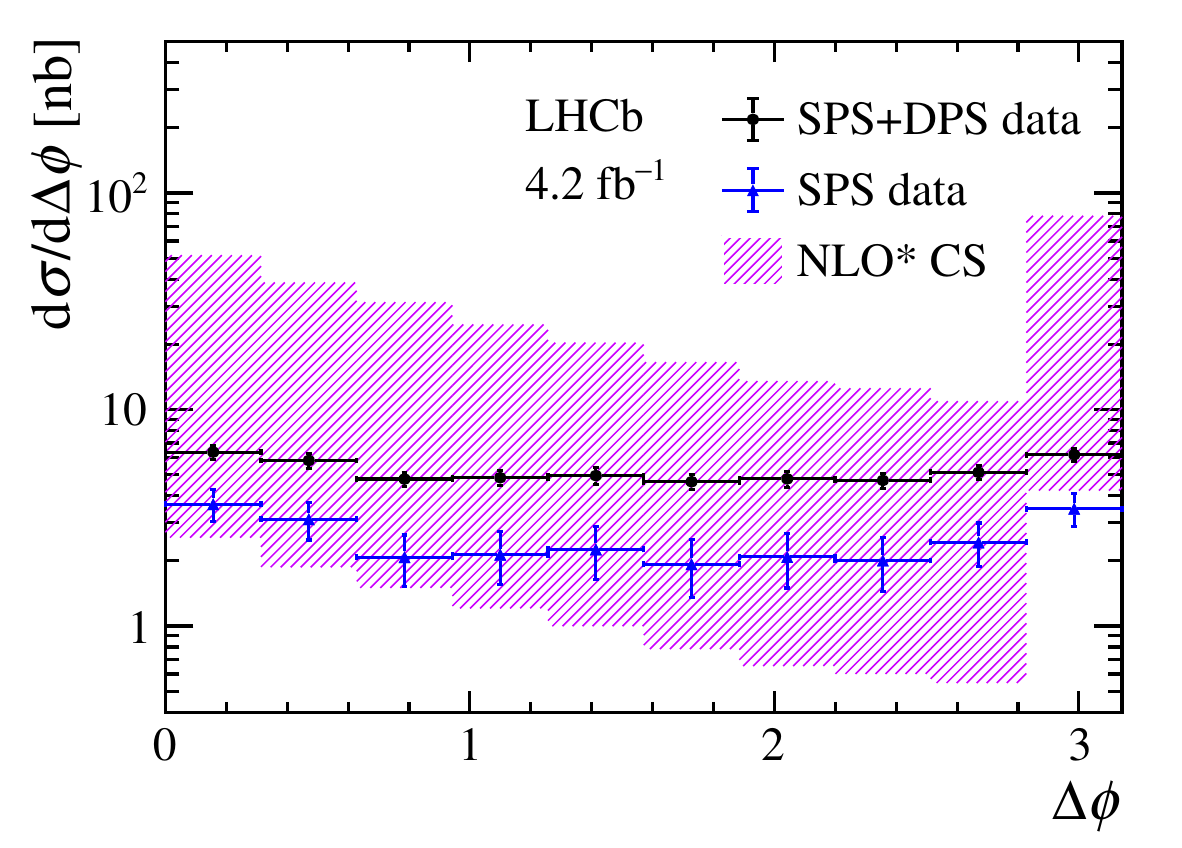}\put(-150,110){(b)}
    \\
    \includegraphics[width=0.41\linewidth]{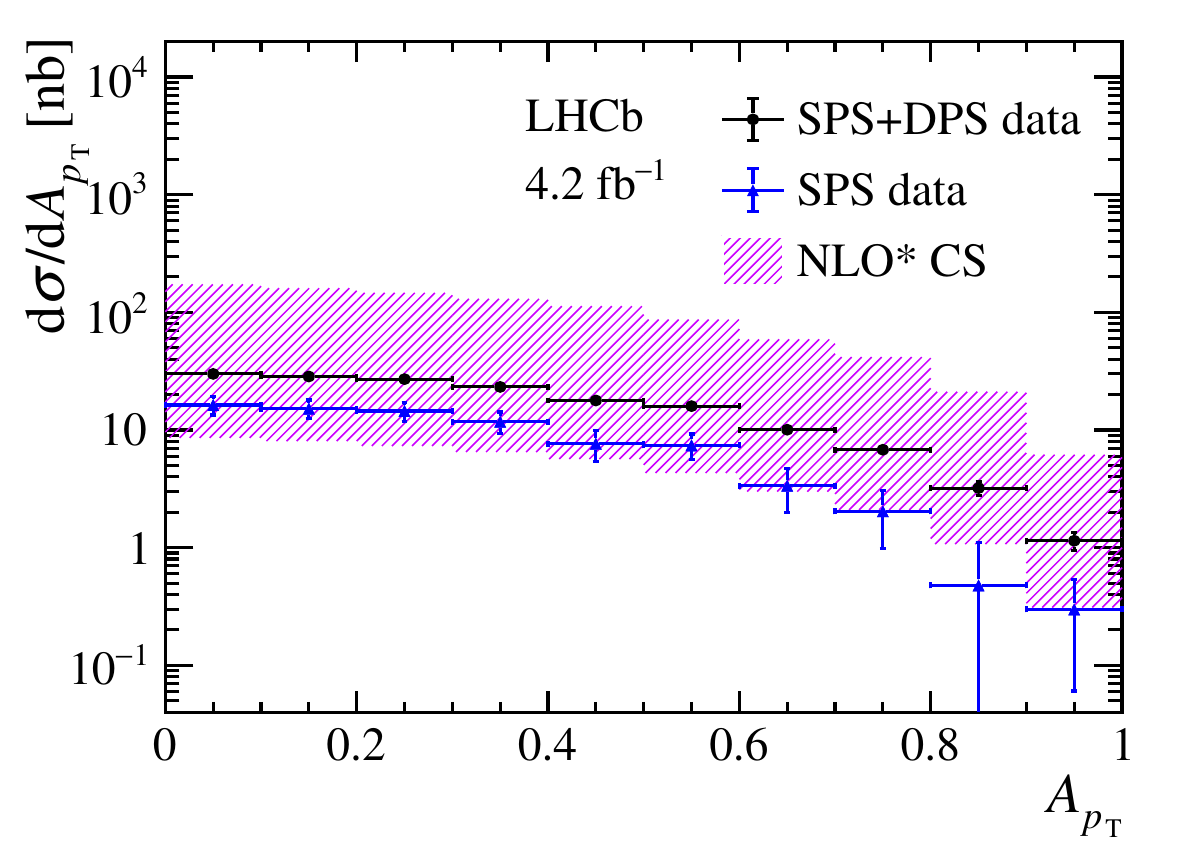}\put(-150,110){(c)}
    \includegraphics[width=0.41\linewidth]{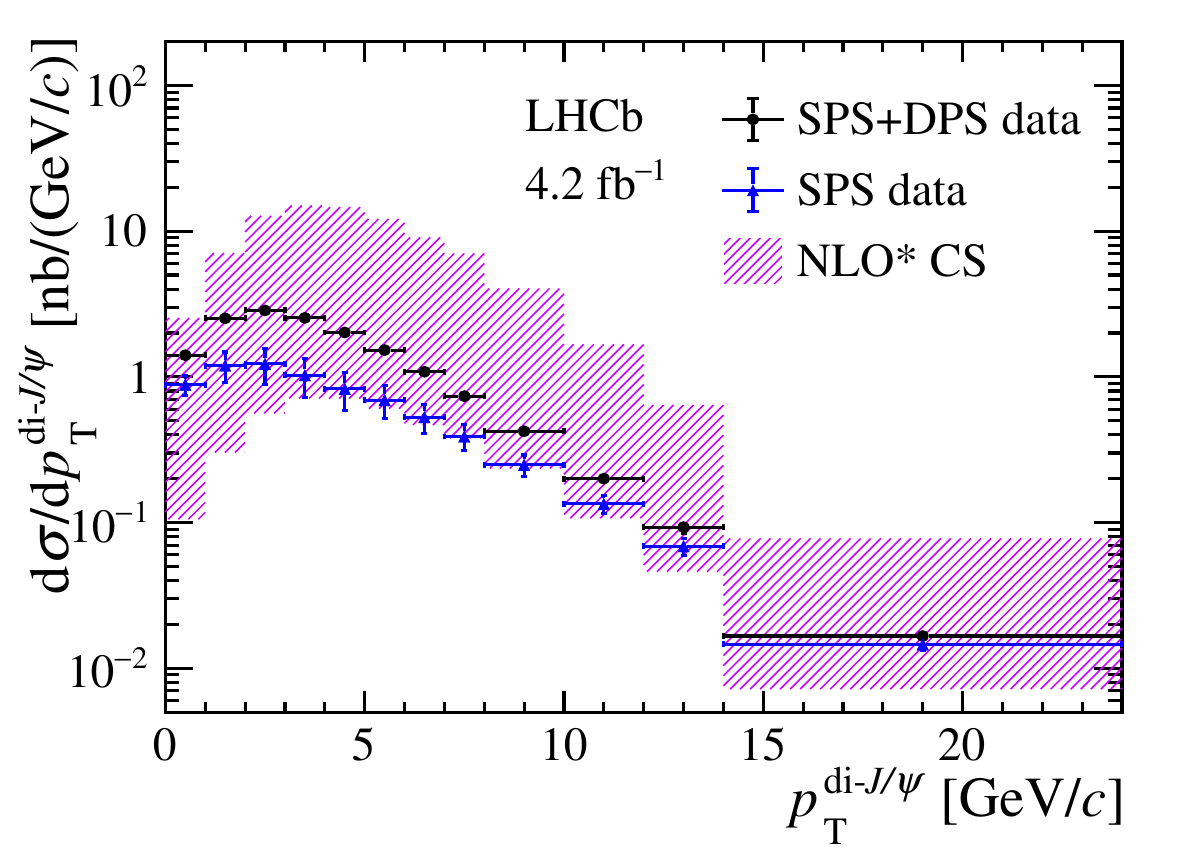}\put(-150,110){(d)}
    \\
    \includegraphics[width=0.41\linewidth]{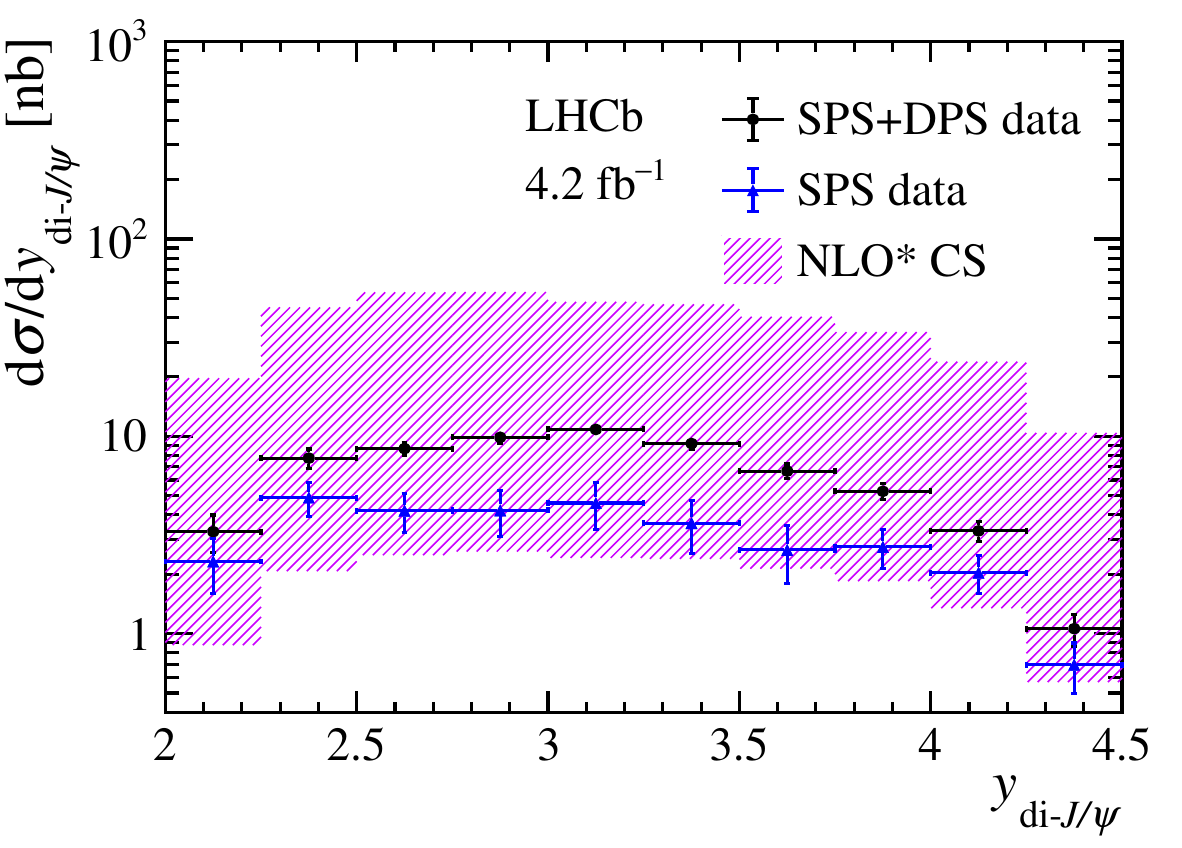}\put(-150,110){(e)}
    \includegraphics[width=0.41\linewidth]{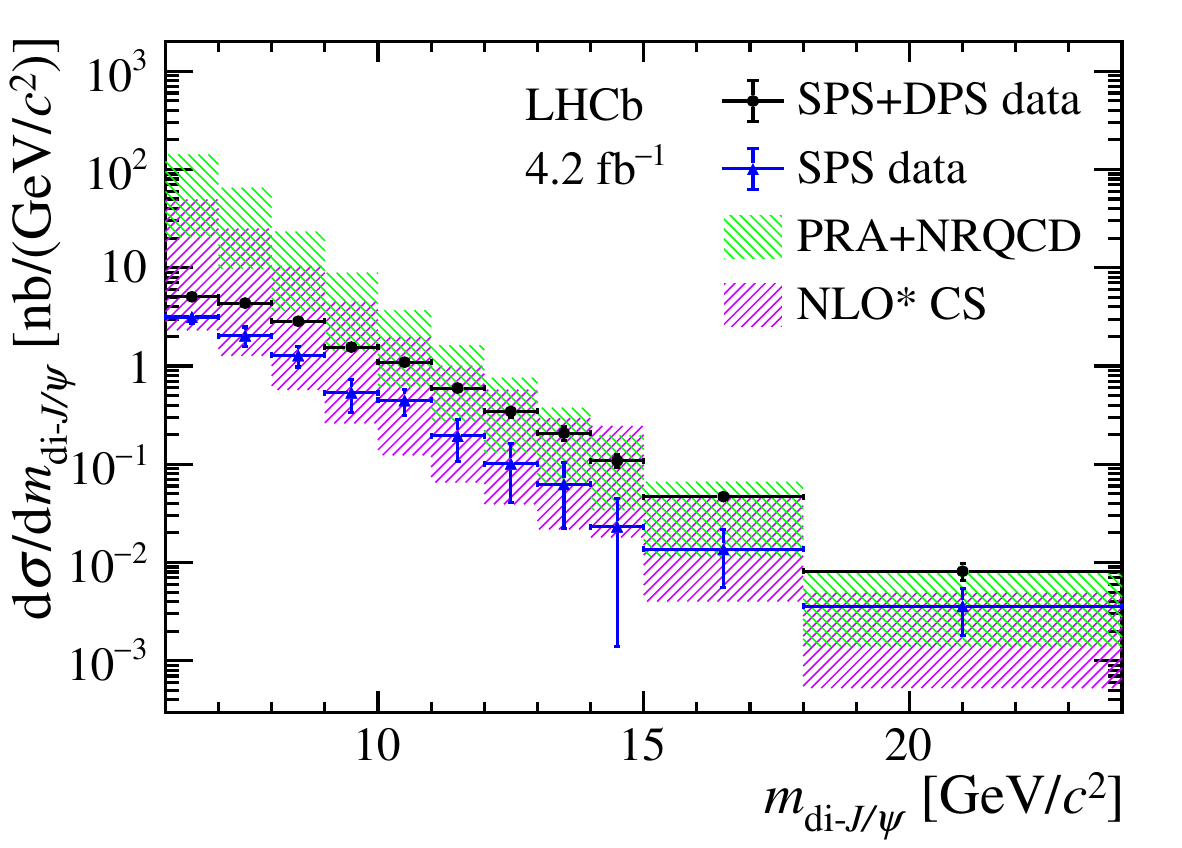}\put(-150,110){(f)}
    \\
    \includegraphics[width=0.41\linewidth]{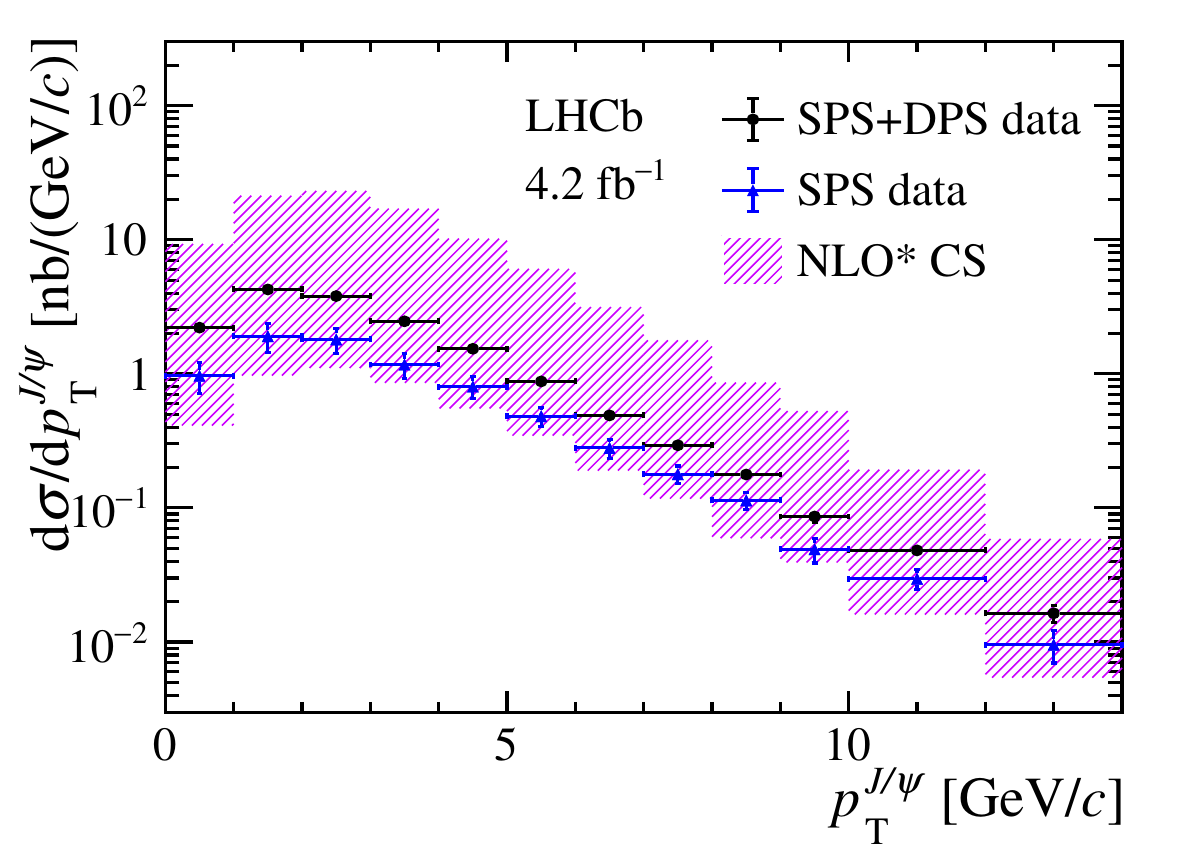}\put(-150,110){(g)}
    \includegraphics[width=0.41\linewidth]{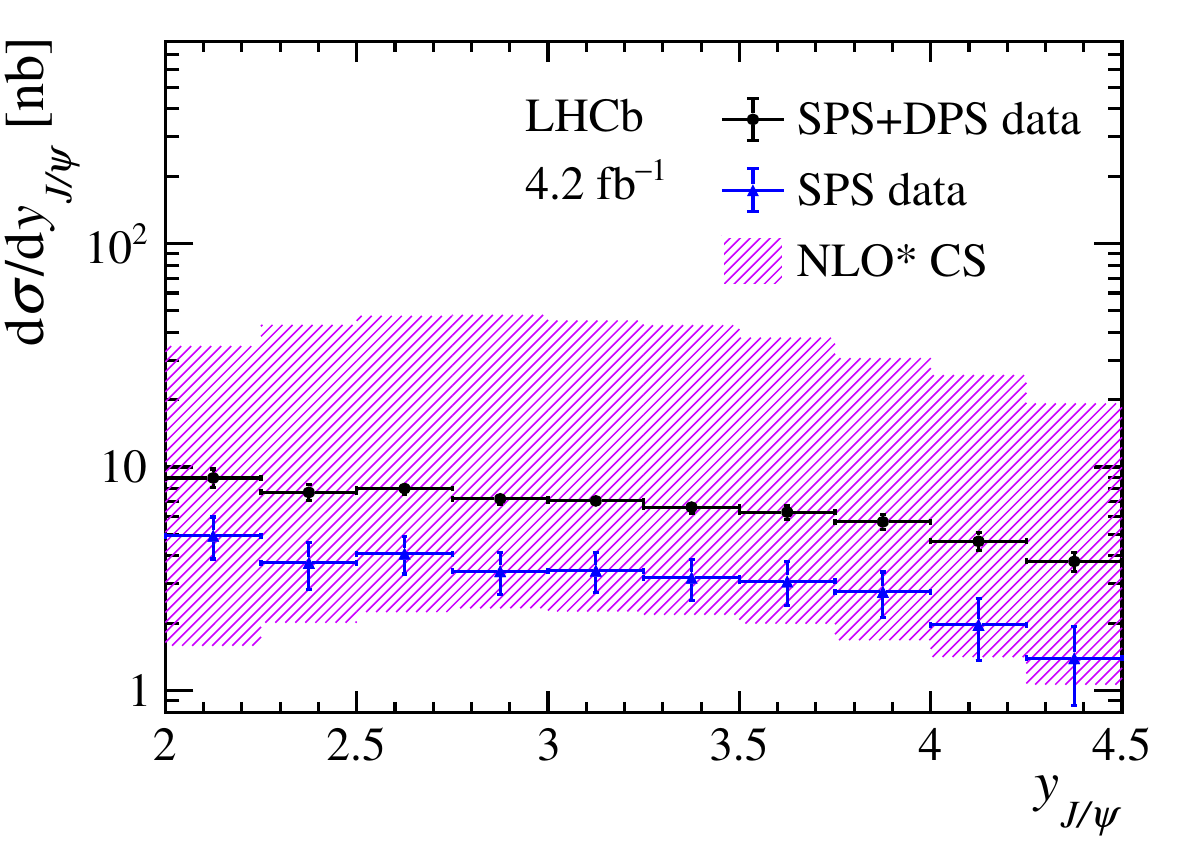}\put(-150,110){(h)}
  \caption{Differential cross-section of di-\jpsi production for SPS+DPS and SPS as a function of (a) $\Delta y$, (b) $\Delta\phi$, (c) $\mathcal{A}_{\pt}$, (d) $\pt^{\text{di-}\jpsi}$, (e) $y_{\text{di-}\jpsi}$, (f) $m_{\text{di-}\jpsi}$, (g) $\pt^{\jpsi}$ and (h) $y_{\jpsi}$, compared with the NLO*~CS predictions for SPS~\cite{Lansberg:2013qka,Shao:2012iz,Shao:2015vga}. The $m_{\text{di-}\jpsi}$ spectrum is also compared with the PRA+NRQCD predictions for SPS~\cite{He:2019qqr}.}
  \label{fig:comTheoryDiJpsi}
\end{figure}
\begin{figure}[tb]
    \centering
    \includegraphics[width=0.6\linewidth]{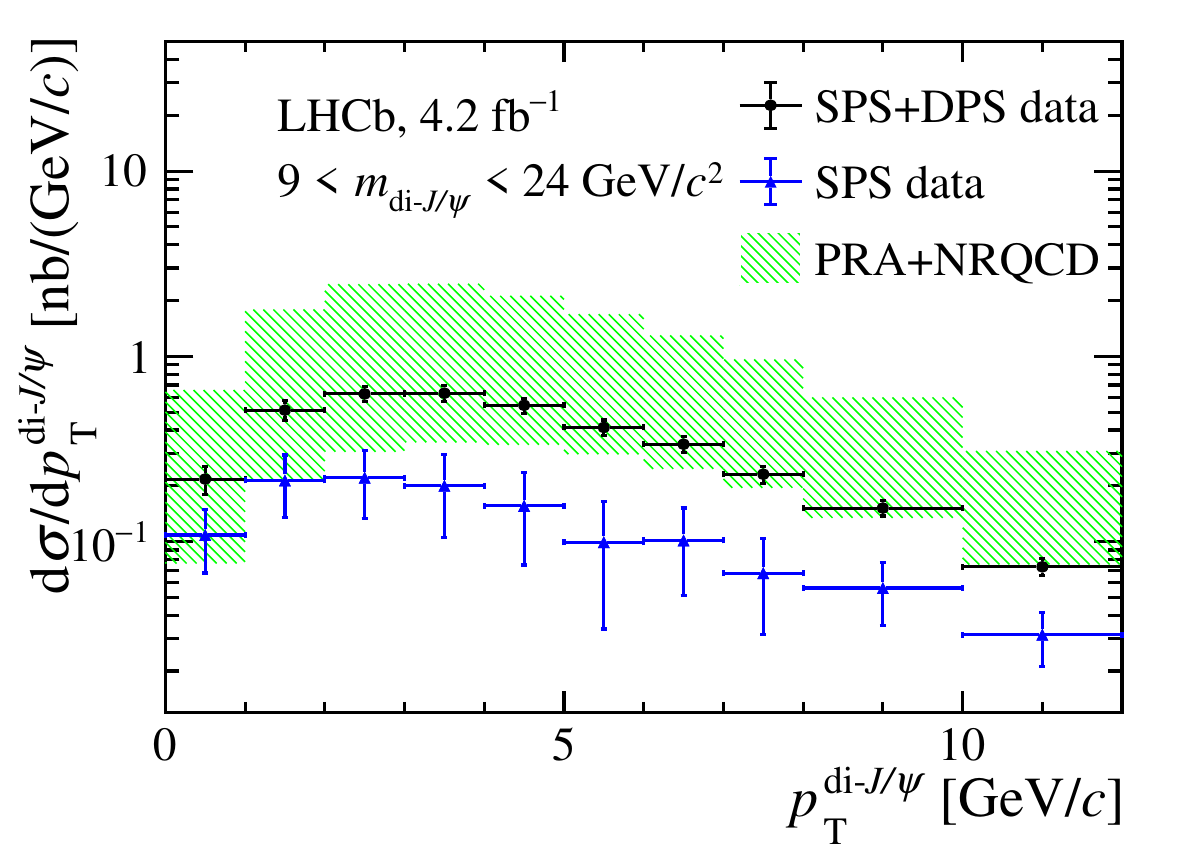}
    \caption{Differential cross-section of di-\jpsi production for SPS+DPS and SPS as a function of $\pt^{\text{di-}\jpsi}$ for candidates with $9<m_{\text{di-}\jpsi}<24\gevcc$, compared with the PRA+NRQCD predictions for SPS~\cite{He:2019qqr}.}
    \label{fig:csPT_largeM}
\end{figure}

The no-loop next-to-leading-order colour-singlet (NLO*~CS) predictions~\cite{Lansberg:2013qka} of the di-\jpsi production via SPS are obtained from HELAC-Onia~\cite{Shao:2012iz,Shao:2015vga}, an automatic matrix element generator for heavy quarkonium physics.
The measured SPS differential cross-sections are compared with the NLO*~CS predictions, as shown in Fig.~\ref{fig:comTheoryDiJpsi}.
The NLO*~CS predictions include the uncertainties from the factorisation and renormalisation scales and subleading PDF uncertainties, correlated between the intervals.
The measurements are consistent with the NLO*~CS calculations within the large theoretical uncertainties.
The mass spectrum is also compared with the predictions combining parton Reggeization approach (PRA)~\cite{Karpishkov:2017kph} and NRQCD factorisation~\cite{He:2019qqr}, which includes a subset of higher-order QCD corrections without ad-hoc kinematic cuts.
Only the renormalisation and factorisation scale uncertainties are considered in the predictions.
In the large $m_{\text{di-}\jpsi}$ region, where the NRQCD-based calculations are well justified, there is a good agreement with the data, while the same predictions exceed the SPS data at small $m_{\text{di-}\jpsi}$, as shown in Fig.~\ref{fig:comTheoryDiJpsi}(f).
The PRA+NRQCD calculations on the $\pt^{\text{di-}\jpsi}$ spectrum are thus further compared with that of the SPS data with $9<m_{\text{di-}\jpsi}<24\gevcc$,
where the PRA+NRACD approach is expected to be validated, as shown in Fig.~\ref{fig:csPT_largeM}.
The predictions are consistent with the SPS cross-sections at small $\pt^{\text{di-}\jpsi}$, but they overestimate the SPS data at $\pt^{\text{di-}\jpsi}$ larger than 3\gevc.

\section{Study of gluon TMDs}
\label{sec:gluonTMD}
\begin{figure}[tb]
    \centering
    \includegraphics[width=0.49\linewidth]{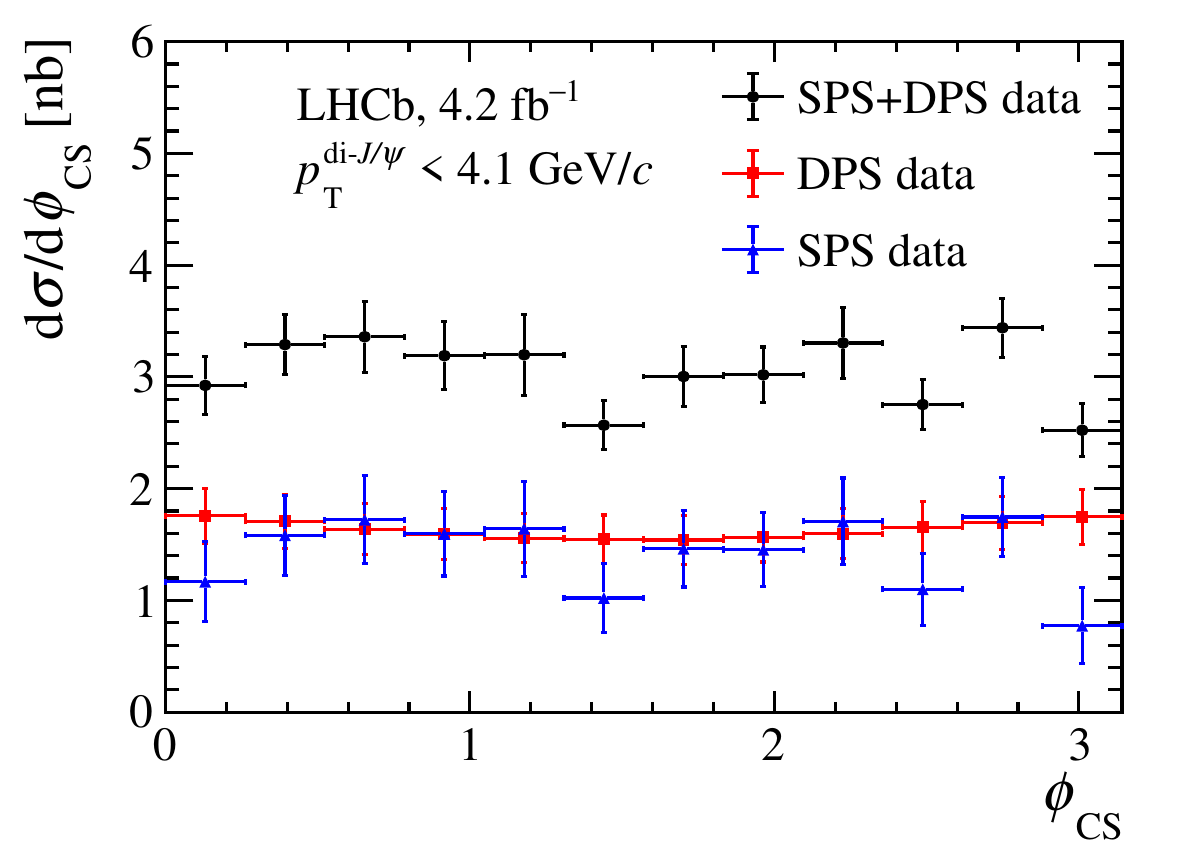}\put(-183,133){(a)}
    \hfil
    \includegraphics[width=0.49\linewidth]{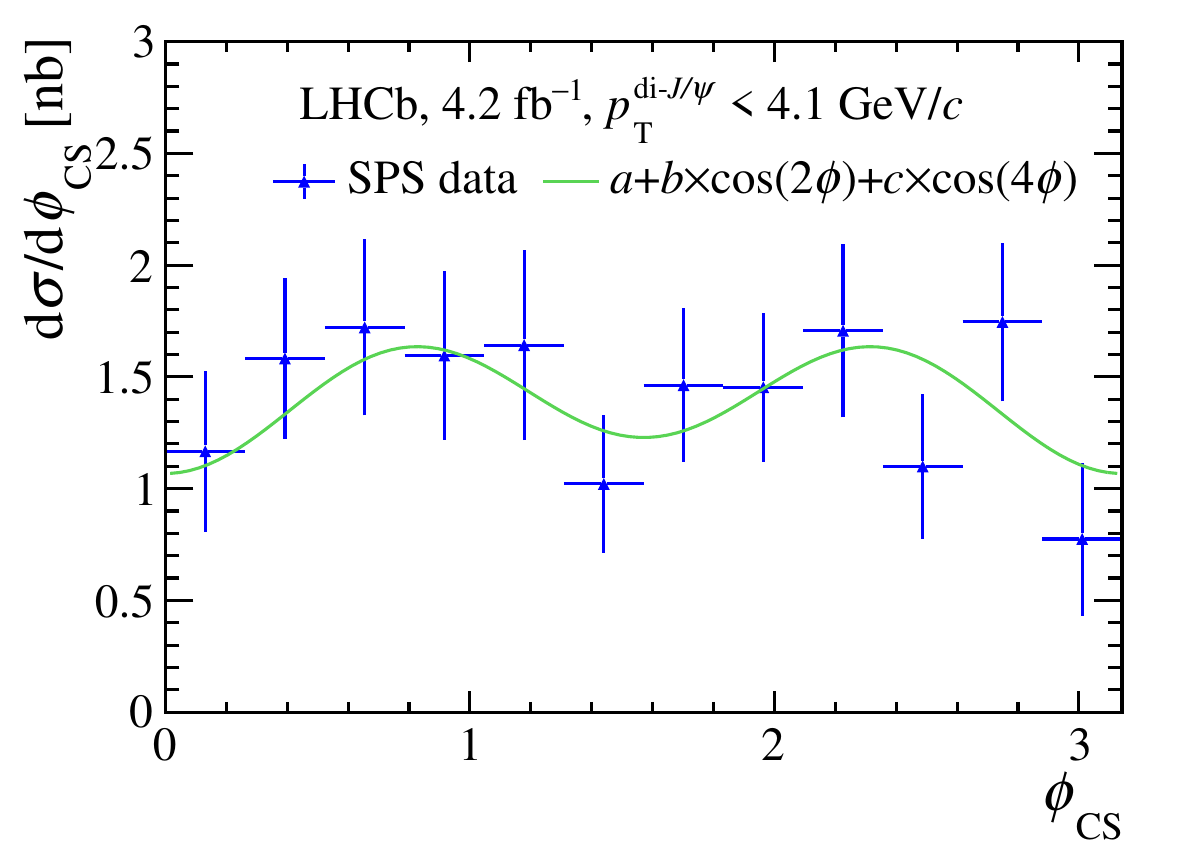}\put(-183,133){(b)}
    \caption{Distribution of $\phi_{\text{CS}}$ (a) with the SPS and DPS contributions separated in the TMD region $\pt^{\text{di-}\jpsi}<4.1\gevc$ and (b) for SPS with the function described in the text overlaid. The systematic uncertainties correlated between intervals are excluded from the error bars.}
    \label{fig:csPHIcs}
\end{figure}

The gluon TMD $h_1^{\perp g}(x,k_{\rm T}^{2},\mu)$ inside unpolarised protons, representing the distribution of linearly polarised gluons, can be probed through the distribution of the azimuthal angle $\phi_{\text{CS}}$ of either \jpsi meson in the Collins-Soper frame.
This frame is the rest frame of the \jpsi pair with the polar axis ($z$-axis) bisecting the angle between the momentum of one proton and the reverse of the momentum of the other proton, the $y$-axis defined to be perpendicular to the plane spanned by the momenta of two protons, and the $x$-axis defined to complete a right-handed Cartesian coordinate system~\cite{Collins:1977iv}.
The prediction of the differential di-\jpsi production cross-section as a function of $\phi_{\text{CS}}$ through SPS is proportional to $a+b\times\cos(2\phi_{\text{CS}})+c\times\cos(4\phi_{\text{CS}})$.
The parameters $a$, $b$ and $c$ encode information on the gluon TMDs as
\begin{align}
a &= F_1 \mathcal{C}[f_1^g f_1^g] + F_2 \mathcal{C}[w_2 h_1^{\perp g} h_1^{\perp g}], \\
b &= F_3\mathcal{C}[w_3f_1^gh_1^{\perp g}]+F_3^\prime\mathcal{C}[w_3^\prime h_1^{\perp g}f_1^g],\\
c &= F_4\mathcal{C}[w_4h_1^{\perp g}h_1^{\perp g}],
\end{align}
where $F_i(^\prime)$ are hard-scattering coefficients,
$w_i(^\prime)$ are the TMD weights common to all gluon-fusion processes originating from unpolarised proton collisions,
and $\mathcal{C}$ denotes the TMD convolutions~\cite{Lansberg:2017dzg,Scarpa:2019fol}.
The calculation is valid in the TMD region with ${\pt^{\text{di-\jpsi}}<\langle m_{\text{di-\jpsi}}\rangle/2}$~\cite{Lansberg:2017dzg,Scarpa:2019fol}.
In this analysis, the $\phi_{\text{CS}}$ distribution is measured in the TMD region ${\pt^{\text{di-}\jpsi}<4.1\gevc}$, since the average value of $m_{\text{di-}\jpsi}$ in the whole fiducial range is $\langle m_{\text{di-}\jpsi}\rangle=8.2\gevcc$.
The measured $\phi_{\text{CS}}$ distributions with the SPS and DPS contributions separated are shown in Fig.~\ref{fig:csPHIcs}(a).
The expectation values $\langle\cos2\phi_{\text{CS}}\rangle$ and $\langle\cos4\phi_{\text{CS}}\rangle$ correspond to half of the ratio of the $\cos n\phi_{\text{CS}}$-modulations present in the TMD cross-section regarding its $\phi_{\text{CS}}$-independent component~\cite{Scarpa:2019fol}, \ie 
$\langle\cos2\phi_{\text{CS}}\rangle = b/2a$ and $\langle\cos4\phi_{\text{CS}}\rangle = c/2a$.
They are calculated as
\begin{align}
\label{eq:calcos2phi}
    \langle\cos2\phi_{\text{CS}}\rangle&=\frac{\sum_i \frac{\deriv\sigma}{\deriv\phi_{\text{CS}}}\big|_i{\Delta\phi_{\text{CS}}}_i\cos2{\phi_{\text{CS}}}_i}{\sum_i \frac{\deriv\sigma}{\deriv\phi_{\text{CS}}}\big|_i{\Delta\phi_{\text{CS}}}_i}, \\
\label{eq:calcos4phi}
    \langle\cos4\phi_{\text{CS}}\rangle&=\frac{\sum_i \frac{\deriv\sigma}{\deriv\phi_{\text{CS}}}\big|_i{\Delta\phi_{\text{CS}}}_i\cos4{\phi_{\text{CS}}}_i}{\sum_i \frac{\deriv\sigma}{\deriv\phi_{\text{CS}}}\big|_i{\Delta\phi_{\text{CS}}}_i},
\end{align}
where the index $i$ denotes each interval, ${\Delta\phi_{\text{CS}}}_i$ is the interval width and ${\phi_{\text{CS}}}_i$ is the interval centre.
The results of $\langle\cos2\phi_{\text{CS}}\rangle$ and $\langle\cos4\phi_{\text{CS}}\rangle$ extracted from the $\phi_{\text{CS}}$ distribution for SPS are
\begin{align*}
    \langle\cos2\phi_{\text{CS}}\rangle= -0.029\pm0.050~(\text{stat})\pm0.009~(\text{syst}), \\
    \langle\cos4\phi_{\text{CS}}\rangle= -0.087\pm0.052~(\text{stat})\pm0.013~(\text{syst}),
\end{align*}
dominated by statistical uncertainties.
The corresponding $\phi_{\text{CS}}$ function given by $a+b\times\cos(2\phi_{\text{CS}})+c\times\cos(4\phi_{\text{CS}})$ is overlaid on the SPS result in Fig.~\ref{fig:csPHIcs}(b).
Its coefficients are fixed to the values calculated by Eqs.~\ref{eq:calcos2phi} and ~\ref{eq:calcos4phi}, and the normalisation is fixed to that of the SPS measurement.
The results are consistent with zero, but the presence of an azimuthal asymmetry at a few percent level is allowed.
The prediction of $\langle\cos2\phi_{\text{CS}}\rangle$ varies from 0.009 to 0.016 due to nonperturbative uncertainties~\cite{Scarpa:2019fol},
also consistent with the measured result given the large uncertainty so far.

\begin{figure}[tb]
	\centering
	\includegraphics[width=0.6\linewidth]{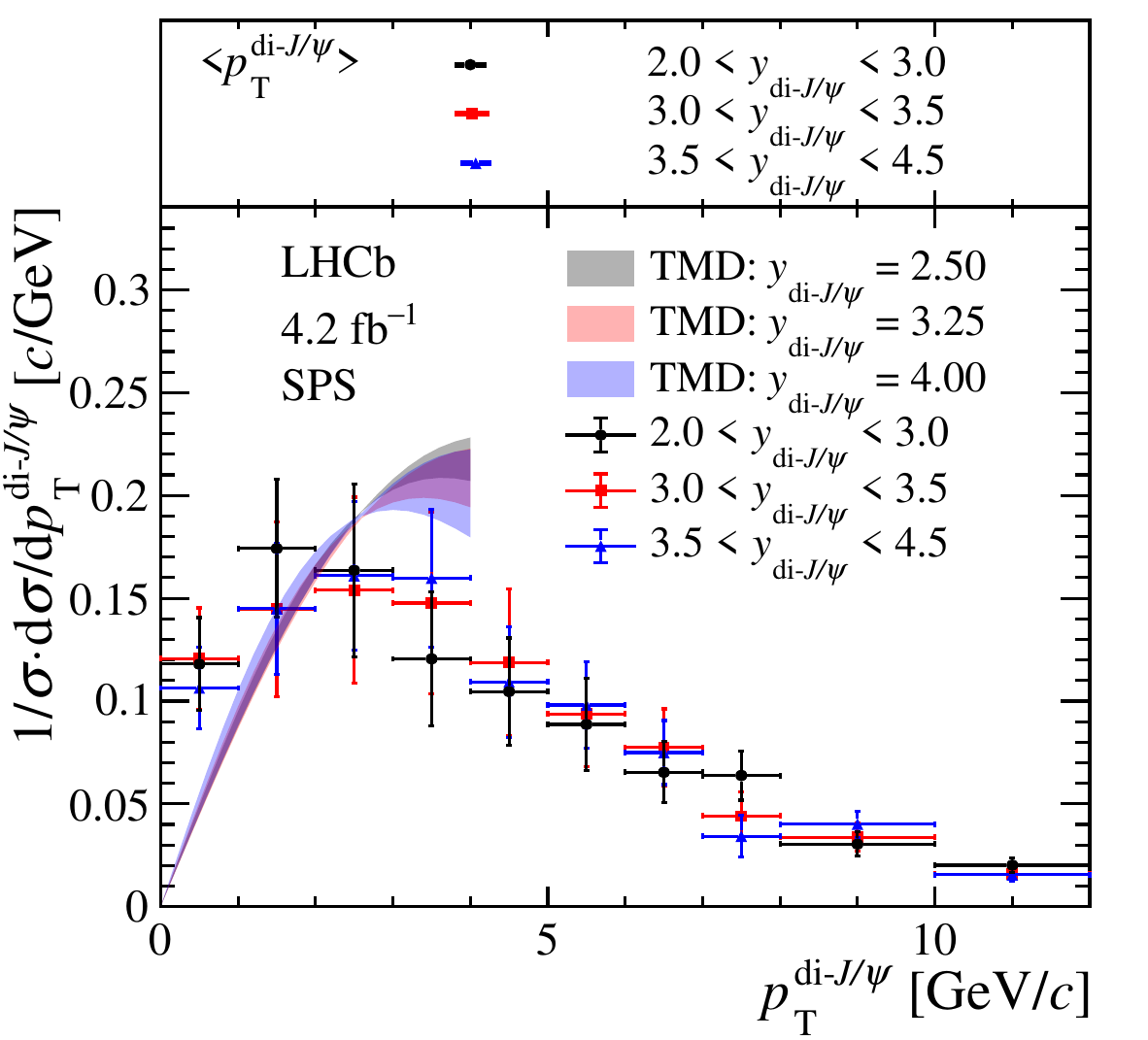}
	\caption{Normalised \pt spectrum of di-\jpsi production in different $y_{\text{di-}\jpsi}$ intervals, compared with TMD predictions~\cite{Scarpa:2019fol} in the TMD region $\pt^{\text{di-}\jpsi}<\langle m_{\text{di-}\jpsi}\rangle/2$. The average values of the $\pt^{\text{di-}\jpsi}$ distributions in three $y_{\text{di-}\jpsi}$ intervals are presented at the top of the figure.}
	\label{fig:diffNormSPSPTY}
\end{figure}
\begin{figure}[tb]
	\centering
	\includegraphics[width=0.6\linewidth]{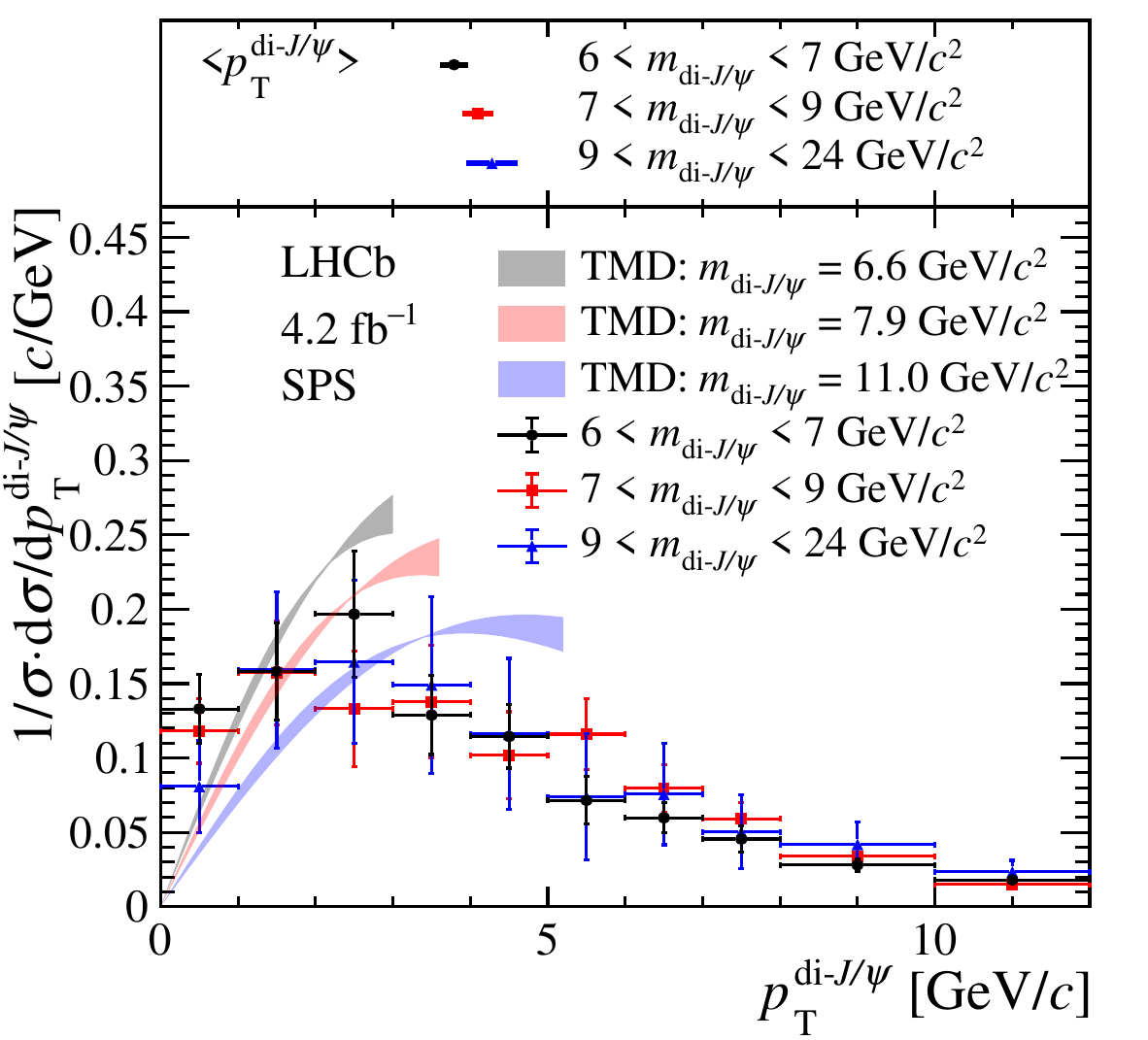}
	\caption{Normalised \pt spectrum of di-\jpsi production in three $m_{\text{di-}\jpsi}$ intervals with $\langle m_{\text{di-}\jpsi}\rangle=$6.6, 7.9 and 11.0\gevcc, compared with TMD predictions~\cite{Scarpa:2019fol} in the TMD region $\pt^{\text{di-}\jpsi}<\langle m_{\text{di-}\jpsi}\rangle/2$. The average values of the $\pt^{\text{di-}\jpsi}$ distributions in three $m_{\text{di-}\jpsi}$ intervals are presented at the top of the figure.}
	\label{fig:diffNormSPSMPT}
\end{figure}
The \pt spectrum of the di-\jpsi signals from SPS can also be used to probe the gluon TMDs, especially $f_1^{g}(x,k_{\rm T}^{2},\mu)$~\cite{Lansberg:2017dzg,Scarpa:2019fol}.
It was pointed out in Ref.~\cite{Scarpa:2019fol} that the variation of the momentum fractions of the two interacting gluons, ${x_{1,2}=m_{\text{di-}\jpsi}e^{\pm y_{\text{di-}\jpsi}}/\sqs}$, do not have significant impact on the shape of the $\pt^{\text{di-\jpsi}}$ spectrum.
The $\pt^{\text{di-}\jpsi}$ spectrum is thus measured in three different intervals of $y_{\text{di-}\jpsi}$ for the SPS process, and the cross-section results are listed in Tables~\ref{table:csDoubleJpsi2DPTY} and~\ref{table:csDoubleJpsi2DSPSPTY} in Appendix~\ref{sec:Tables} for SPS+DPS and SPS separately.
The distributions are normalised for comparison in Fig.~\ref{fig:diffNormSPSPTY}.
They are consistent with each other within the uncertainties.
The average values of the $\pt^{\text{di-}\jpsi}$ distributions in three $y_{\text{di-}\jpsi}$ intervals are also presented at the top of Fig.~\ref{fig:diffNormSPSPTY}, and show no significant variations.
The TMD predictions~\cite{Scarpa:2019fol}, which are only applicable in the TMD region $\pt^{\text{di-}\jpsi}<\langle m_{\text{di-}\jpsi}\rangle/2$, are also shown in Fig.~\ref{fig:diffNormSPSPTY},
and peak at higher $\pt^{\text{di-\jpsi}}$ than the measured distributions.

In addition, the study of the dependence of TMDs on the renormalisation and rapidity scales, requires a measurement of the \pt spectrum at different $m_{\text{di-}\jpsi}$~\cite{Scarpa:2019fol}.
The differential cross-sections $\deriv\sigma/\deriv\pt^{\text{di-}\jpsi}$ in the three intervals \mbox{$6<m_{\text{di-}\jpsi}<7\gevcc$}, \mbox{$7<m_{\text{di-}\jpsi}<9\gevcc$} and \mbox{$9<m_{\text{di-}\jpsi}<24\gevcc$}, are listed in Tables~\ref{table:csDoubleJpsi2DMPT} and~\ref{table:csDoubleJpsi2DSPSMPT} in Appendix~\ref{sec:Tables} for SPS+DPS and SPS separately.
The normalised \pt spectra of the di-\jpsi production for SPS in different $m_{\text{di-}\jpsi}$ intervals with the expected values of $\langle m_{\text{di-}\jpsi}\rangle=6.6$, 7.9 and 11.0\gevcc, respectively, are compared in Figure~\ref{fig:diffNormSPSMPT},
with the TMD predictions~\cite{Scarpa:2019fol} overlaid in the TMD region.
According to the prediction, the \pt spectrum would broaden as $m_{\text{di-}\jpsi}$ increases~\cite{Scarpa:2019fol},
but no obvious broadening of the \pt spectrum can be seen in the TMD region due to the large uncertainties.
The average values of the $\pt^{\text{di-}\jpsi}$ distributions in three $m_{\text{di-}\jpsi}$ intervals are also presented at the top of Fig.~\ref{fig:diffNormSPSMPT},
and slightly increase with mass.

\section{Conclusion}
The \jpsi-pair production cross-section in $pp$ collisions at $\sqs=13\tev$ is measured to be $16.36\pm0.28~(\text{stat})\pm0.88~(\text{syst})\nb$ using a data sample corresponding to an integrated luminosity of $4.2\invfb$ collected by the LHCb experiment,
with both \jpsi mesons in the range of $\pt<14\gevc$ and $2.0<y<4.5$.
The contributions from DPS and SPS are separated based on distinctive $\Delta y$ dependences of their corresponding cross-sections.
The effective cross-section characterising the DPS process is determined to be $\sigma_{\text{eff}}=13.1\pm1.8~(\text{stat})\pm2.3~(\text{syst})\mbarn$,
and is consistent with most of the existing measurements.
The differential cross-sections in SPS are consistent with the NLO*~CS predictions which are plagued by large theoretical uncertainties.
The cross-sections predicted by PRA+NRQCD overshoot the SPS data at small $m_{\text{di-}\jpsi}$ and agree with them at large $m_{\text{di-}\jpsi}$.

The gluon TMDs are probed via the $\phi_{\text{CS}}$ distribution and the $\pt^{\text{di-}\jpsi}$ spectrum from the SPS process.
The extracted values of $\langle\cos2\phi_{\text{CS}}\rangle$ and $\langle\cos4\phi_{\text{CS}}\rangle$ are consistent with zero, but the presence of an azimuthal asymmetry at a few percent level is allowed.
The $\pt^{\text{di-}\jpsi}$ spectra are consistent with each other in different $y_{\text{di-}\jpsi}$ intervals, and peak at lower $\pt^{\text{di-}\jpsi}$ values than the theoretical predictions.
No significant broadening of the \pt spectrum with increasing $m_{\text{di-}\jpsi}$ is seen in the TMD region, but the average value of the $\pt^{\text{di-}\jpsi}$ distribution increases slightly with $m_{\text{di-}\jpsi}$.
The results provide important experimental inputs to study gluon TMDs and to improve theoretical models.
The ongoing data taking by the LHCb experiment will enable more precise measurements of production cross-sections, and allow to test the predictions of gluon TMDs with higher precision.

\section*{Acknowledgements}
\noindent We would like to thank J.-P. Lansberg, H.-S. Shao, Y.-Q. Ma and L.-P. Sun for the helpful discussions on the quarkonium and quarkonium-pair production,
and J.-P. Lansberg, H.-S. Shao, D. Boer, J. Bor, A. Colpani Serri, Z.-G. He and M. A. Nefedov for providing the SPS calculations.
We express our gratitude to our colleagues in the CERN
accelerator departments for the excellent performance of the LHC. We
thank the technical and administrative staff at the LHCb
institutes.
We acknowledge support from CERN and from the national agencies:
CAPES, CNPq, FAPERJ and FINEP (Brazil); 
MOST and NSFC (China); 
CNRS/IN2P3 (France); 
BMBF, DFG and MPG (Germany); 
INFN (Italy); 
NWO (Netherlands); 
MNiSW and NCN (Poland); 
MCID/IFA (Romania); 
MICINN (Spain); 
SNSF and SER (Switzerland); 
NASU (Ukraine); 
STFC (United Kingdom); 
DOE NP and NSF (USA).
We acknowledge the computing resources that are provided by CERN, IN2P3
(France), KIT and DESY (Germany), INFN (Italy), SURF (Netherlands),
PIC (Spain), GridPP (United Kingdom), 
CSCS (Switzerland), IFIN-HH (Romania), CBPF (Brazil),
and Polish WLCG (Poland).
We are indebted to the communities behind the multiple open-source
software packages on which we depend.
Individual groups or members have received support from
ARC and ARDC (Australia);
Key Research Program of Frontier Sciences of CAS, CAS PIFI, CAS CCEPP, 
Fundamental Research Funds for the Central Universities, 
and Sci. \& Tech. Program of Guangzhou (China);
Minciencias (Colombia);
EPLANET, Marie Sk\l{}odowska-Curie Actions, ERC and NextGenerationEU (European Union);
A*MIDEX, ANR, IPhU and Labex P2IO, and R\'{e}gion Auvergne-Rh\^{o}ne-Alpes (France);
AvH Foundation (Germany);
ICSC (Italy); 
GVA, XuntaGal, GENCAT, Inditex, InTalent and Prog.~Atracci\'on Talento, CM (Spain);
SRC (Sweden);
the Leverhulme Trust, the Royal Society
 and UKRI (United Kingdom).

\clearpage

\newcommand{\xx}{\ensuremath{\kern 0.5em }}
\newcommand{\xxx}{\ensuremath{\kern 0.75em }}

\appendix

\section{Differential cross-sections}
\label{sec:Tables}

The measured differential cross-sections of di-\jpsi production as functions of $\Delta y$, $\Delta\phi$, $\mathcal{A}_{\pt}$, $\pt^{\text{di-}\jpsi}$, $y_{\text{di-}\jpsi}$, $m_{\text{di-}\jpsi}$, $\pt^{\jpsi}$ and $y_{\jpsi}$ are listed in Tables~\ref{table:csDoubleJpsiDY}--\ref{table:csDoubleJpsiJpsiY}.
The differential cross-sections of di-\jpsi production for SPS process are listed in Tables~\ref{table:csDoubleJpsiDYSPS}--\ref{table:csDoubleJpsiJpsiYSPS}.
The differential cross-sections $\deriv\sigma/\deriv\pt^{\text{di-}\jpsi}$ in three different intervals of $y_{\text{di-}\jpsi}$ are listed in Tables~\ref{table:csDoubleJpsi2DPTY} and~\ref{table:csDoubleJpsi2DSPSPTY} for SPS+DPS and SPS separately.
The differential cross-sections $\deriv\sigma/\deriv\pt^{\text{di-}\jpsi}$ in three intervals of $m_{\text{di-}\jpsi}$ are listed in Tables~\ref{table:csDoubleJpsi2DMPT} and~\ref{table:csDoubleJpsi2DSPSMPT} for SPS+DPS and SPS separately.

\begin{table}[h]
\centering
\caption{Differential cross-sections $\deriv\sigma/\deriv\Delta y$ of di-\jpsi production. The first uncertainties are statistical, and the second systematic.}
\label{table:csDoubleJpsiDY}
\begin{tabular}{cc}
\hline $\Delta y$ & $\deriv\sigma/\deriv\Delta y$ [nb] \\\hline
0.0\,--\,0.2& $19.46\pm 0.87\pm 1.15$ \\
0.2\,--\,0.4& $14.88\pm 0.56\pm 0.81$ \\
0.4\,--\,0.6& $12.57\pm 0.50\pm 0.69$ \\
0.6\,--\,0.8& $\xx9.32\pm 0.37\pm 0.52$ \\
0.8\,--\,1.0& $\xx7.29\pm 0.33\pm 0.41$ \\
1.0\,--\,1.2& $\xx5.52\pm 0.28\pm 0.31$ \\
1.2\,--\,1.4& $\xx4.36\pm 0.27\pm 0.24$ \\
1.4\,--\,1.6& $\xx3.34\pm 0.26\pm 0.21$ \\
1.6\,--\,1.8& $\xx2.42\pm 0.26\pm 0.15$ \\
1.8\,--\,2.0& $\xx1.67\pm 0.23\pm 0.18$ \\
2.0\,--\,2.5& $\xx0.50\pm 0.13\pm 0.05$ \\
\hline
\end{tabular}
\end{table}

\begin{table}[]
\centering
\label{table:csDoubleJpsiDPHI}
\caption{Differential cross-sections $\deriv\sigma/\deriv\Delta\phi$ of di-\jpsi production. The first uncertainties are statistical, and the second systematic.}
\begin{tabular}{cc}
\hline $\Delta\phi/\pi$ & $\deriv\sigma/\deriv\Delta\phi$ [nb] \\\hline
0.0\,--\,0.1& $ 6.35\pm 0.30\pm 0.37$ \\
0.1\,--\,0.2& $ 5.81\pm 0.31\pm 0.34$ \\
0.2\,--\,0.3& $ 4.77\pm 0.24\pm 0.27$ \\
0.3\,--\,0.4& $ 4.84\pm 0.30\pm 0.27$ \\
0.4\,--\,0.5& $ 4.95\pm 0.35\pm 0.29$ \\
0.5\,--\,0.6& $ 4.64\pm 0.27\pm 0.27$ \\
0.6\,--\,0.7& $ 4.78\pm 0.30\pm 0.27$ \\
0.7\,--\,0.8& $ 4.70\pm 0.25\pm 0.27$ \\
0.8\,--\,0.9& $ 5.12\pm 0.23\pm 0.29$ \\
0.9\,--\,1.0& $ 6.18\pm 0.29\pm 0.35$ \\
\hline
\end{tabular}
\end{table}

\begin{table}[]
\centering
\caption{Differential cross-sections $\deriv\sigma/\deriv\mathcal{A}_{\pt}$ of di-\jpsi production. The first uncertainties are statistical, and the second systematic.}
\label{table:csDoubleJpsiAPT}
\begin{tabular}{cc}
\hline $\mathcal{A}_{\pt}$ & $\deriv\sigma/\deriv\mathcal{A}_{\pt}$ [nb] \\\hline
0.0\,--\,0.1& $29.98\pm 1.17\pm 1.68$ \\
0.1\,--\,0.2& $28.50\pm 1.04\pm 1.59$ \\
0.2\,--\,0.3& $27.06\pm 1.34\pm 1.52$ \\
0.3\,--\,0.4& $23.23\pm 1.06\pm 1.30$ \\
0.4\,--\,0.5& $17.82\pm 1.04\pm 1.10$ \\
0.5\,--\,0.6& $15.95\pm 0.80\pm 0.98$ \\
0.6\,--\,0.7& $10.06\pm 0.64\pm 0.56$ \\
0.7\,--\,0.8& $\xx6.79\pm 0.59\pm 0.37$ \\
0.8\,--\,0.9& $\xx3.21\pm 0.38\pm 0.21$ \\
0.9\,--\,1.0& $\xx1.14\pm 0.18\pm 0.08$ \\
\hline
\end{tabular}
\end{table}

\begin{table}[]
\centering
\caption{Differential cross-sections $\deriv\sigma/\deriv\pt^{\text{di-}\jpsi}$ of di-\jpsi production. The first uncertainties are statistical, and the second systematic.}
\label{table:csDoubleJpsiPT}
\begin{tabular}{cc}
\hline $\pt^{\text{di-}\jpsi}$ [\ensuremath{\text{Ge\kern -0.1em V\!/}c}] & $\deriv\sigma/\deriv \pt^{\text{di-}\jpsi}$ [nb/(\ensuremath{\text{Ge\kern -0.1em V\!/}c})] \\\hline
0\,--\,1& $ 1.408\pm 0.089\pm 0.083$ \\
1\,--\,2& $ 2.523\pm 0.126\pm 0.150$ \\
2\,--\,3& $ 2.858\pm 0.158\pm 0.164$ \\
3\,--\,4& $ 2.542\pm 0.110\pm 0.146$ \\
4\,--\,5& $ 2.017\pm 0.081\pm 0.120$ \\
5\,--\,6& $ 1.527\pm 0.073\pm 0.091$ \\
6\,--\,7& $ 1.085\pm 0.048\pm 0.065$ \\
7\,--\,8& $ 0.738\pm 0.038\pm 0.044$ \\
8\,--\,10& $ 0.424\pm 0.018\pm 0.027$ \\
10\,--\,12& $ 0.200\pm 0.011\pm 0.013$ \\
12\,--\,14& $ 0.093\pm 0.007\pm 0.005$ \\
14\,--\,24& $ 0.017\pm 0.001\pm 0.001$ \\
\hline
\end{tabular}
\end{table}

\begin{table}[]
\centering
\caption{Differential cross-sections $\deriv\sigma/\deriv y_{\text{di-}\jpsi}$ of di-\jpsi production. The first uncertainties are statistical, and the second systematic.}
\label{table:csDoubleJpsiY}
\begin{tabular}{cc}
\hline $y_{\text{di-}\jpsi}$ & $\deriv\sigma/\deriv y_{\text{di-}\jpsi}$ [nb] \\\hline
2.00\,--\,2.25& $\xx3.28\pm 0.64\pm 0.29$ \\
2.25\,--\,2.50& $\xx7.75\pm 0.53\pm 0.69$ \\
2.50\,--\,2.75& $\xx8.66\pm 0.36\pm 0.52$ \\
2.75\,--\,3.00& $\xx9.88\pm 0.32\pm 0.59$ \\
3.00\,--\,3.25& $10.83\pm 0.32\pm 0.62$ \\
3.25\,--\,3.50& $\xx9.18\pm 0.30\pm 0.54$ \\
3.50\,--\,3.75& $\xx6.66\pm 0.21\pm 0.54$ \\
3.75\,--\,4.00& $\xx5.26\pm 0.19\pm 0.44$ \\
4.00\,--\,4.25& $\xx3.31\pm 0.19\pm 0.35$ \\
4.25\,--\,4.50& $\xx1.06\pm 0.16\pm 0.12$ \\
\hline
\end{tabular}
\end{table}

\begin{table}[]
\centering
\caption{Differential cross-sections $\deriv\sigma/\deriv m_{\text{di-}\jpsi}$ of di-\jpsi production. The first uncertainties are statistical, and the second systematic.}
\label{table:csDoubleJpsiM}
\begin{tabular}{cc}
\hline $m_{\text{di-}\jpsi}$ [\ensuremath{\text{Ge\kern -0.1em V\!/}c}] & $\deriv\sigma/\deriv m_{\text{di-}\jpsi}$ [nb/(\ensuremath{\text{Ge\kern -0.1em V\!/}c})] \\\hline
6\,--\,7& $ 5.059\pm 0.200\pm 0.283$ \\
7\,--\,8& $ 4.364\pm 0.131\pm 0.244$ \\
8\,--\,9& $ 2.851\pm 0.092\pm 0.155$ \\
9\,--\,10& $ 1.555\pm 0.067\pm 0.085$ \\
10\,--\,11& $ 1.092\pm 0.062\pm 0.062$ \\
11\,--\,12& $ 0.595\pm 0.049\pm 0.034$ \\
12\,--\,13& $ 0.344\pm 0.037\pm 0.026$ \\
13\,--\,14& $ 0.207\pm 0.029\pm 0.016$ \\
14\,--\,15& $ 0.108\pm 0.015\pm 0.008$ \\
15\,--\,18& $ 0.047\pm 0.005\pm 0.003$ \\
18\,--\,24& $ 0.008\pm 0.002\pm 0.001$ \\
\hline
\end{tabular}
\end{table}

\begin{table}[]
\centering
\caption{Differential cross-sections $\deriv\sigma/\deriv\pt^{\jpsi}$ of di-\jpsi production. The first uncertainties are statistical, and the second systematic.}
\label{table:csDoubleJpsiJpsiPT}
\begin{tabular}{cc}
\hline $\pt^{\jpsi}$ [\ensuremath{\text{Ge\kern -0.1em V\!/}c}] & $\deriv\sigma/\deriv \pt^{\jpsi}$ [nb/(\ensuremath{\text{Ge\kern -0.1em V\!/}c})] \\\hline
0\,--\,1& $ 2.209\pm 0.093\pm 0.132$ \\
1\,--\,2& $ 4.255\pm 0.124\pm 0.255$ \\
2\,--\,3& $ 3.798\pm 0.090\pm 0.208$ \\
3\,--\,4& $ 2.464\pm 0.062\pm 0.135$ \\
4\,--\,5& $ 1.536\pm 0.044\pm 0.085$ \\
5\,--\,6& $ 0.878\pm 0.027\pm 0.048$ \\
6\,--\,7& $ 0.489\pm 0.019\pm 0.028$ \\
7\,--\,8& $ 0.293\pm 0.013\pm 0.016$ \\
8\,--\,9& $ 0.177\pm 0.010\pm 0.011$ \\
9\,--\,10& $ 0.086\pm 0.007\pm 0.005$ \\
10\,--\,12& $ 0.048\pm 0.003\pm 0.003$ \\
12\,--\,14& $ 0.016\pm 0.002\pm 0.001$ \\
\hline
\end{tabular}
\end{table}

\begin{table}[]
\centering
\caption{Differential cross-sections $\deriv\sigma/\deriv y_{\jpsi}$ of di-\jpsi production. The first uncertainties are statistical, and the second systematic.}
\label{table:csDoubleJpsiJpsiY}
\begin{tabular}{cc}
\hline $y_{\jpsi}$ & $\deriv\sigma/\deriv y_{\jpsi}$ [nb] \\\hline
2.00\,--\,2.25& $ 8.93\pm 0.56\pm 0.65$ \\
2.25\,--\,2.50& $ 7.69\pm 0.29\pm 0.56$ \\
2.50\,--\,2.75& $ 8.01\pm 0.21\pm 0.44$ \\
2.75\,--\,3.00& $ 7.20\pm 0.18\pm 0.39$ \\
3.00\,--\,3.25& $ 7.05\pm 0.16\pm 0.38$ \\
3.25\,--\,3.50& $ 6.61\pm 0.15\pm 0.36$ \\
3.50\,--\,3.75& $ 6.28\pm 0.15\pm 0.45$ \\
3.75\,--\,4.00& $ 5.68\pm 0.15\pm 0.41$ \\
4.00\,--\,4.25& $ 4.65\pm 0.16\pm 0.39$ \\
4.25\,--\,4.50& $ 3.78\pm 0.20\pm 0.32$ \\
\hline
\end{tabular}
\end{table}

\begin{table}[]
\centering
\caption{Differential cross-sections $\deriv\sigma/\deriv\Delta y$ of di-\jpsi production in SPS. The first uncertainties are statistical, and the second systematic.}
\label{table:csDoubleJpsiDYSPS}
\begin{tabular}{cc}
\hline $\Delta y$ & $\deriv\sigma/\deriv\Delta y$ [nb] \\\hline
0.0\,--\,0.2& $\kern 0.25em 12.76\pm 1.29\pm 1.13$ \\
0.2\,--\,0.4& $\xxx8.72\pm 1.03\pm 0.85$ \\
0.4\,--\,0.6& $\xxx6.97\pm 0.94\pm 0.74$ \\
0.6\,--\,0.8& $\xxx4.32\pm 0.80\pm 0.62$ \\
0.8\,--\,1.0& $\xxx2.89\pm 0.70\pm 0.53$ \\
1.0\,--\,1.2& $\xxx1.74\pm 0.60\pm 0.43$ \\
1.2\,--\,1.4& $\xxx1.15\pm 0.53\pm 0.37$ \\
1.4\,--\,1.6& $\xxx0.75\pm 0.45\pm 0.31$ \\
1.6\,--\,1.8& $\xxx0.37\pm 0.39\pm 0.24$ \\
1.8\,--\,2.0& $\xxx0.22\pm 0.31\pm 0.22$ \\
2.0\,--\,2.5& $-0.07\pm 0.16\pm 0.08$ \\
\hline
\end{tabular}
\end{table}

\begin{table}[]
\centering
\caption{Differential cross-sections $\deriv\sigma/\deriv\Delta\phi$ of di-\jpsi production in SPS. The first uncertainties are statistical, and the second systematic.}
\label{table:csDoubleJpsiDPHISPS}
\begin{tabular}{cc}
\hline $\Delta\phi/\pi$ & $\deriv\sigma/\deriv\Delta\phi$ [nb] \\\hline
0.0\,--\,0.1& $ 3.65\pm 0.48\pm 0.39$ \\
0.1\,--\,0.2& $ 3.11\pm 0.49\pm 0.37$ \\
0.2\,--\,0.3& $ 2.08\pm 0.45\pm 0.33$ \\
0.3\,--\,0.4& $ 2.14\pm 0.48\pm 0.33$ \\
0.4\,--\,0.5& $ 2.26\pm 0.52\pm 0.34$ \\
0.5\,--\,0.6& $ 1.93\pm 0.47\pm 0.33$ \\
0.6\,--\,0.7& $ 2.09\pm 0.48\pm 0.33$ \\
0.7\,--\,0.8& $ 2.01\pm 0.46\pm 0.33$ \\
0.8\,--\,0.9& $ 2.43\pm 0.44\pm 0.34$ \\
0.9\,--\,1.0& $ 3.48\pm 0.48\pm 0.37$ \\
\hline
\end{tabular}
\end{table}

\begin{table}[]
\centering
\caption{Differential cross-sections $\deriv\sigma/\deriv\mathcal{A}_{\pt}$ of di-\jpsi production in SPS. The first uncertainties are statistical, and the second systematic.}
\label{table:csDoubleJpsiAPTSPS}
\begin{tabular}{cc}
\hline $\mathcal{A}_{\pt}$ & $\deriv\sigma/\deriv\mathcal{A}_{\pt}$ [nb] \\\hline
0.0\,--\,0.1& $16.31\pm 2.26\pm 1.83$ \\
0.1\,--\,0.2& $15.19\pm 2.15\pm 1.76$ \\
0.2\,--\,0.3& $14.50\pm 2.22\pm 1.67$ \\
0.3\,--\,0.4& $11.78\pm 1.93\pm 1.48$ \\
0.4\,--\,0.5& $\xx7.65\pm 1.77\pm 1.32$ \\
0.5\,--\,0.6& $\xx7.41\pm 1.44\pm 1.14$ \\
0.6\,--\,0.7& $\xx3.37\pm 1.14\pm 0.77$ \\
0.7\,--\,0.8& $\xx2.04\pm 0.90\pm 0.54$ \\
0.8\,--\,0.9& $\xx0.48\pm 0.54\pm 0.32$ \\
0.9\,--\,1.0& $\xx0.30\pm 0.21\pm 0.10$ \\
\hline
\end{tabular}
\end{table}

\begin{table}[]
\centering
\caption{Differential cross-sections $\deriv\sigma/\deriv\pt^{\text{di-}\jpsi}$ of di-\jpsi production in SPS. The first uncertainties are statistical, and the second systematic.}
\label{table:csDoubleJpsiPTSPS}
\begin{tabular}{cc}
\hline $\pt^{\text{di-}\jpsi}$ [\ensuremath{\text{Ge\kern -0.1em V\!/}c}] & $\deriv\sigma/\deriv \pt^{\text{di-}\jpsi}$ [nb/(\ensuremath{\text{Ge\kern -0.1em V\!/}c})] \\\hline
0\,--\,1& $ 0.886\pm 0.116\pm 0.084$ \\
1\,--\,2& $ 1.196\pm 0.225\pm 0.173$ \\
2\,--\,3& $ 1.229\pm 0.279\pm 0.201$ \\
3\,--\,4& $ 1.026\pm 0.241\pm 0.184$ \\
4\,--\,5& $ 0.827\pm 0.186\pm 0.149$ \\
5\,--\,6& $ 0.693\pm 0.139\pm 0.107$ \\
6\,--\,7& $ 0.529\pm 0.092\pm 0.074$ \\
7\,--\,8& $ 0.390\pm 0.062\pm 0.048$ \\
8\,--\,10& $ 0.249\pm 0.031\pm 0.028$ \\
10\,--\,12& $ 0.134\pm 0.014\pm 0.013$ \\
12\,--\,14& $ 0.069\pm 0.008\pm 0.005$ \\
14\,--\,24& $ 0.015\pm 0.001\pm 0.001$ \\
\hline
\end{tabular}
\end{table}

\begin{table}[]
\centering
\caption{Differential cross-sections $\deriv\sigma/\deriv y_{\text{di-}\jpsi}$ of di-\jpsi production in SPS. The first uncertainties are statistical, and the second systematic.}
\label{table:csDoubleJpsiYSPS}
\begin{tabular}{cc}
\hline $y_{\text{di-}\jpsi}$ & $\deriv\sigma/\deriv y_{\text{di-}\jpsi}$ [nb] \\\hline
2.00\,--\,2.25& $ 2.32\pm 0.65\pm 0.29$ \\
2.25\,--\,2.50& $ 4.89\pm 0.67\pm 0.69$ \\
2.50\,--\,2.75& $ 4.20\pm 0.72\pm 0.59$ \\
2.75\,--\,3.00& $ 4.21\pm 0.86\pm 0.72$ \\
3.00\,--\,3.25& $ 4.59\pm 0.94\pm 0.77$ \\
3.25\,--\,3.50& $ 3.63\pm 0.84\pm 0.68$ \\
3.50\,--\,3.75& $ 2.66\pm 0.60\pm 0.62$ \\
3.75\,--\,4.00& $ 2.75\pm 0.40\pm 0.46$ \\
4.00\,--\,4.25& $ 2.04\pm 0.26\pm 0.36$ \\
4.25\,--\,4.50& $ 0.70\pm 0.16\pm 0.12$ \\
\hline
\end{tabular}
\end{table}

\begin{table}[]
\centering
\caption{Differential cross-sections $\deriv\sigma/\deriv m_{\text{di-}\jpsi}$ of di-\jpsi production in SPS. The first uncertainties are statistical, and the second systematic.}
\label{table:csDoubleJpsiMSPS}
\begin{tabular}{cc}
\hline $m_{\text{di-}\jpsi}$ [\ensuremath{\text{Ge\kern -0.1em V\!/}c}] & $\deriv\sigma/\deriv m_{\text{di-}\jpsi}$ [nb/(\ensuremath{\text{Ge\kern -0.1em V\!/}c})] \\\hline
6\,--\,7& $ 3.147\pm 0.336\pm 0.286$ \\
7\,--\,8& $ 2.039\pm 0.353\pm 0.290$ \\
8\,--\,9& $ 1.283\pm 0.240\pm 0.190$ \\
9\,--\,10& $ 0.535\pm 0.159\pm 0.117$ \\
10\,--\,11& $ 0.446\pm 0.110\pm 0.078$ \\
11\,--\,12& $ 0.194\pm 0.075\pm 0.047$ \\
12\,--\,13& $ 0.101\pm 0.050\pm 0.033$ \\
13\,--\,14& $ 0.063\pm 0.035\pm 0.020$ \\
14\,--\,15& $ 0.023\pm 0.019\pm 0.011$ \\
15\,--\,18& $ 0.014\pm 0.007\pm 0.004$ \\
18\,--\,24& $ 0.004\pm 0.002\pm 0.001$ \\
\hline
\end{tabular}
\end{table}

\begin{table}[]
\centering
\caption{Differential cross-sections $\deriv\sigma/\deriv\pt^{\jpsi}$ of di-\jpsi production in SPS. The first uncertainties are statistical, and the second systematic.}
\label{table:csDoubleJpsiJpsiPTSPS}
\begin{tabular}{cc}
\hline $\pt^{\jpsi}$ [\ensuremath{\text{Ge\kern -0.1em V\!/}c}] & $\deriv\sigma/\deriv \pt^{\jpsi}$ [nb/(\ensuremath{\text{Ge\kern -0.1em V\!/}c})] \\\hline
0\,--\,1& $ 0.966\pm 0.198\pm 0.159$ \\
1\,--\,2& $ 1.907\pm 0.354\pm 0.303$ \\
2\,--\,3& $ 1.807\pm 0.295\pm 0.246$ \\
3\,--\,4& $ 1.173\pm 0.192\pm 0.160$ \\
4\,--\,5& $ 0.804\pm 0.112\pm 0.095$ \\
5\,--\,6& $ 0.484\pm 0.062\pm 0.052$ \\
6\,--\,7& $ 0.280\pm 0.035\pm 0.029$ \\
7\,--\,8& $ 0.178\pm 0.021\pm 0.017$ \\
8\,--\,9& $ 0.114\pm 0.014\pm 0.011$ \\
9\,--\,10& $ 0.049\pm 0.009\pm 0.006$ \\
10\,--\,12& $ 0.030\pm 0.004\pm 0.003$ \\
12\,--\,14& $ 0.010\pm 0.002\pm 0.001$ \\
\hline
\end{tabular}
\end{table}

\begin{table}[]
\centering
\caption{Differential cross-sections $\deriv\sigma/\deriv y_{\jpsi}$ of di-\jpsi production in SPS. The first uncertainties are statistical, and the second systematic.}
\label{table:csDoubleJpsiJpsiYSPS}
\begin{tabular}{cc}
\hline $y_{\jpsi}$ & $\deriv\sigma/\deriv y_{\jpsi}$ [nb] \\\hline
2.00\,--\,2.25& $ 4.92\pm 0.79\pm 0.68$ \\
2.25\,--\,2.50& $ 3.72\pm 0.63\pm 0.61$ \\
2.50\,--\,2.75& $ 4.09\pm 0.59\pm 0.50$ \\
2.75\,--\,3.00& $ 3.42\pm 0.56\pm 0.47$ \\
3.00\,--\,3.25& $ 3.44\pm 0.53\pm 0.44$ \\
3.25\,--\,3.50& $ 3.19\pm 0.51\pm 0.43$ \\
3.50\,--\,3.75& $ 3.08\pm 0.48\pm 0.49$ \\
3.75\,--\,4.00& $ 2.76\pm 0.44\pm 0.45$ \\
4.00\,--\,4.25& $ 1.97\pm 0.41\pm 0.44$ \\
4.25\,--\,4.50& $ 1.39\pm 0.39\pm 0.37$ \\
\hline
\end{tabular}
\end{table}

\begin{table}[]
\centering
\caption{Differential cross-sections $\deriv\sigma/\deriv\pt^{\text{di-}\jpsi}$ [nb/(\ensuremath{\text{Ge\kern -0.1em V\!/}c})] of di-\jpsi production in intervals of $y_{\text{di-}\jpsi}$. The first uncertainties are statistical, and the second systematic.}
\label{table:csDoubleJpsi2DPTY}
\begin{tabular}{cccc}
\hline $\pt^{\text{di-}\jpsi} [\ensuremath{\text{Ge\kern -0.1em V\!/}c}]$ & $2.0<y_{\text{di-}\jpsi}<3.0$ & $3.0<y_{\text{di-}\jpsi}<3.5$ & $3.5<y_{\text{di-}\jpsi}<4.5$ \\\hline
0\,--\,1& $ 0.626\pm 0.072\pm 0.037$ & $ 0.426\pm 0.035\pm 0.025$ & $ 0.334\pm 0.029\pm 0.020$ \\
1\,--\,2& $ 1.140\pm 0.103\pm 0.068$ & $ 0.763\pm 0.055\pm 0.045$ & $ 0.611\pm 0.037\pm 0.036$ \\
2\,--\,3& $ 1.237\pm 0.137\pm 0.071$ & $ 0.881\pm 0.046\pm 0.050$ & $ 0.707\pm 0.040\pm 0.040$ \\
3\,--\,4& $ 1.054\pm 0.084\pm 0.060$ & $ 0.823\pm 0.051\pm 0.047$ & $ 0.665\pm 0.038\pm 0.038$ \\
4\,--\,5& $ 0.874\pm 0.063\pm 0.052$ & $ 0.648\pm 0.038\pm 0.039$ & $ 0.486\pm 0.030\pm 0.029$ \\
5\,--\,6& $ 0.679\pm 0.060\pm 0.040$ & $ 0.469\pm 0.027\pm 0.028$ & $ 0.375\pm 0.026\pm 0.022$ \\
6\,--\,7& $ 0.483\pm 0.036\pm 0.029$ & $ 0.339\pm 0.021\pm 0.020$ & $ 0.263\pm 0.020\pm 0.016$ \\
7\,--\,8& $ 0.390\pm 0.031\pm 0.023$ & $ 0.201\pm 0.014\pm 0.012$ & $ 0.139\pm 0.014\pm 0.008$ \\
8\,--\,10& $ 0.192\pm 0.014\pm 0.012$ & $ 0.123\pm 0.007\pm 0.008$ & $ 0.112\pm 0.008\pm 0.007$ \\
10\,--\,12& $ 0.105\pm 0.009\pm 0.007$ & $ 0.052\pm 0.004\pm 0.003$ & $ 0.042\pm 0.005\pm 0.003$ \\
\hline
\end{tabular}

\end{table}

\begin{table}[]
\centering
\caption{Differential cross-sections $\deriv\sigma/\deriv\pt^{\text{di-}\jpsi}$ [nb/(\ensuremath{\text{Ge\kern -0.1em V\!/}c})] of di-\jpsi production in SPS in intervals of $y_{\text{di-}\jpsi}$. The first uncertainties are statistical, and the second systematic.}
\label{table:csDoubleJpsi2DSPSPTY}
\begin{tabular}{cccc}
\hline $\pt^{\text{di-}\jpsi} [\ensuremath{\text{Ge\kern -0.1em V\!/}c}]$ & $2.0<y_{\text{di-}\jpsi}<3.0$ & $3.0<y_{\text{di-}\jpsi}<3.5$ & $3.5<y_{\text{di-}\jpsi}<4.5$ \\\hline
0\,--\,1& $ 0.428\pm 0.078\pm 0.036$ & $ 0.238\pm 0.044\pm 0.027$ & $ 0.198\pm 0.035\pm 0.021$ \\
1\,--\,2& $ 0.632\pm 0.125\pm 0.072$ & $ 0.286\pm 0.087\pm 0.058$ & $ 0.270\pm 0.061\pm 0.043$ \\
2\,--\,3& $ 0.593\pm 0.164\pm 0.082$ & $ 0.304\pm 0.093\pm 0.068$ & $ 0.299\pm 0.070\pm 0.050$ \\
3\,--\,4& $ 0.437\pm 0.121\pm 0.075$ & $ 0.291\pm 0.091\pm 0.063$ & $ 0.297\pm 0.064\pm 0.046$ \\
4\,--\,5& $ 0.379\pm 0.094\pm 0.063$ & $ 0.234\pm 0.069\pm 0.050$ & $ 0.203\pm 0.050\pm 0.036$ \\
5\,--\,6& $ 0.322\pm 0.078\pm 0.047$ & $ 0.185\pm 0.048\pm 0.035$ & $ 0.182\pm 0.038\pm 0.025$ \\
6\,--\,7& $ 0.237\pm 0.050\pm 0.033$ & $ 0.153\pm 0.034\pm 0.024$ & $ 0.139\pm 0.027\pm 0.017$ \\
7\,--\,8& $ 0.231\pm 0.038\pm 0.024$ & $ 0.087\pm 0.022\pm 0.014$ & $ 0.064\pm 0.017\pm 0.010$ \\
8\,--\,10& $ 0.110\pm 0.018\pm 0.013$ & $ 0.067\pm 0.011\pm 0.009$ & $ 0.075\pm 0.010\pm 0.007$ \\
10\,--\,12& $ 0.073\pm 0.010\pm 0.007$ & $ 0.031\pm 0.005\pm 0.003$ & $ 0.029\pm 0.005\pm 0.003$ \\
\hline
\end{tabular}

\end{table}

\begin{table}[]
\centering
\caption{Differential cross-sections $\deriv\sigma/\deriv\pt^{\text{di-}\jpsi}$ [nb/(\ensuremath{\text{Ge\kern -0.1em V\!/}c})] of di-\jpsi production in intervals of $m_{\text{di-}\jpsi}$ [\ensuremath{\text{Ge\kern -0.1em V\!/}c^2}]. The first uncertainties are statistical, and the second systematic.}
\label{table:csDoubleJpsi2DMPT}
\begin{tabular}{cccc}
\hline $\pt^{\text{di-}\jpsi} [\ensuremath{\text{Ge\kern -0.1em V\!/}c}]$ & $6<m_{\text{di-}\jpsi}<7\gevcc$ & $7<m_{\text{di-}\jpsi}<9\gevcc$ & $9<m_{\text{di-}\jpsi}<24\gevcc$ \\\hline
0\,--\,1& $ 0.563\pm 0.063\pm 0.033$ & $ 0.626\pm 0.053\pm 0.037$ & $ 0.217\pm 0.035\pm 0.013$ \\
1\,--\,2& $ 0.867\pm 0.087\pm 0.051$ & $ 1.138\pm 0.071\pm 0.067$ & $ 0.512\pm 0.056\pm 0.030$ \\
2\,--\,3& $ 1.025\pm 0.132\pm 0.059$ & $ 1.213\pm 0.069\pm 0.069$ & $ 0.627\pm 0.046\pm 0.036$ \\
3\,--\,4& $ 0.743\pm 0.062\pm 0.043$ & $ 1.167\pm 0.074\pm 0.067$ & $ 0.632\pm 0.053\pm 0.036$ \\
4\,--\,5& $ 0.587\pm 0.051\pm 0.035$ & $ 0.883\pm 0.050\pm 0.053$ & $ 0.544\pm 0.039\pm 0.032$ \\
5\,--\,6& $ 0.363\pm 0.041\pm 0.022$ & $ 0.742\pm 0.050\pm 0.044$ & $ 0.413\pm 0.032\pm 0.025$ \\
6\,--\,7& $ 0.264\pm 0.024\pm 0.016$ & $ 0.491\pm 0.032\pm 0.029$ & $ 0.335\pm 0.026\pm 0.020$ \\
7\,--\,8& $ 0.182\pm 0.024\pm 0.011$ & $ 0.327\pm 0.022\pm 0.019$ & $ 0.231\pm 0.020\pm 0.014$ \\
8\,--\,10& $ 0.102\pm 0.010\pm 0.007$ & $ 0.170\pm 0.011\pm 0.011$ & $ 0.152\pm 0.010\pm 0.010$ \\
10\,--\,12& $ 0.059\pm 0.007\pm 0.004$ & $ 0.067\pm 0.006\pm 0.004$ & $ 0.073\pm 0.006\pm 0.005$ \\
\hline
\end{tabular}
\end{table}

\begin{table}[]
\centering
\caption{Differential cross-sections $\deriv\sigma/\deriv\pt^{\text{di-}\jpsi}$ [nb/(\ensuremath{\text{Ge\kern -0.1em V\!/}c})] of di-\jpsi production in SPS in intervals of $m_{\text{di-}\jpsi}$ [\ensuremath{\text{Ge\kern -0.1em V\!/}c^2}]. The first uncertainties are statistical, and the second systematic.}
\label{table:csDoubleJpsi2DSPSMPT}
\begin{tabular}{cccc}
\hline $\pt^{\text{di-}\jpsi} [\ensuremath{\text{Ge\kern -0.1em V\!/}c}]$ & $6<m_{\text{di-}\jpsi}<7\gevcc$ & $7<m_{\text{di-}\jpsi}<9\gevcc$ & $9<m_{\text{di-}\jpsi}<24\gevcc$ \\\hline
0\,--\,1& $ 0.400\pm 0.067\pm 0.032$ & $ 0.376\pm 0.064\pm 0.038$ & $ 0.109\pm 0.038\pm 0.015$ \\
1\,--\,2& $ 0.475\pm 0.103\pm 0.055$ & $ 0.501\pm 0.114\pm 0.081$ & $ 0.214\pm 0.070\pm 0.037$ \\
2\,--\,3& $ 0.591\pm 0.146\pm 0.062$ & $ 0.424\pm 0.131\pm 0.093$ & $ 0.221\pm 0.073\pm 0.048$ \\
3\,--\,4& $ 0.387\pm 0.079\pm 0.047$ & $ 0.438\pm 0.126\pm 0.087$ & $ 0.200\pm 0.081\pm 0.051$ \\
4\,--\,5& $ 0.344\pm 0.062\pm 0.036$ & $ 0.324\pm 0.093\pm 0.068$ & $ 0.156\pm 0.067\pm 0.046$ \\
5\,--\,6& $ 0.215\pm 0.046\pm 0.022$ & $ 0.369\pm 0.073\pm 0.050$ & $ 0.099\pm 0.054\pm 0.037$ \\
6\,--\,7& $ 0.180\pm 0.026\pm 0.015$ & $ 0.253\pm 0.047\pm 0.033$ & $ 0.102\pm 0.042\pm 0.028$ \\
7\,--\,8& $ 0.137\pm 0.024\pm 0.010$ & $ 0.188\pm 0.029\pm 0.020$ & $ 0.068\pm 0.030\pm 0.019$ \\
8\,--\,10& $ 0.084\pm 0.010\pm 0.006$ & $ 0.108\pm 0.014\pm 0.011$ & $ 0.056\pm 0.017\pm 0.012$ \\
10\,--\,12& $ 0.054\pm 0.007\pm 0.004$ & $ 0.048\pm 0.006\pm 0.004$ & $ 0.031\pm 0.009\pm 0.006$ \\
\hline
\end{tabular}
\end{table}

\clearpage

\addcontentsline{toc}{section}{References}
\bibliographystyle{LHCb}
\bibliography{main,standard,LHCb-PAPER,LHCb-CONF,LHCb-DP,LHCb-TDR}

\newpage
\centerline
{\large\bf LHCb collaboration}
\begin
{flushleft}
\small
R.~Aaij$^{35}$\lhcborcid{0000-0003-0533-1952},
A.S.W.~Abdelmotteleb$^{54}$\lhcborcid{0000-0001-7905-0542},
C.~Abellan~Beteta$^{48}$,
F.~Abudin{\'e}n$^{54}$\lhcborcid{0000-0002-6737-3528},
T.~Ackernley$^{58}$\lhcborcid{0000-0002-5951-3498},
B.~Adeva$^{44}$\lhcborcid{0000-0001-9756-3712},
M.~Adinolfi$^{52}$\lhcborcid{0000-0002-1326-1264},
P.~Adlarson$^{78}$\lhcborcid{0000-0001-6280-3851},
H.~Afsharnia$^{11}$,
C.~Agapopoulou$^{46}$\lhcborcid{0000-0002-2368-0147}
% ...

{\footnotesize \it

% ...
$^{11}$Universit{\'e} Clermont Auvergne, CNRS/IN2P3, LPC, Clermont-Ferrand, France\\
% ...
$^{35}$Nikhef National Institute for Subatomic Physics, Amsterdam, Netherlands\\
% ...
$^{44}$Instituto Galego de F{\'\i}sica de Altas Enerx{\'\i}as (IGFAE), Universidade de Santiago de Compostela, Santiago de Compostela, Spain\\
% ...
$^{46}$European Organization for Nuclear Research (CERN), Geneva, Switzerland\\
% ...
$^{48}$Physik-Institut, Universit{\"a}t Z{\"u}rich, Z{\"u}rich, Switzerland\\
% ...
$^{52}$H.H. Wills Physics Laboratory, University of Bristol, Bristol, United Kingdom\\
% ...
$^{54}$Department of Physics, University of Warwick, Coventry, United Kingdom\\
% ...
$^{57}$School of Physics and Astronomy, University of Glasgow, Glasgow, United Kingdom\\
$^{58}$Oliver Lodge Laboratory, University of Liverpool, Liverpool, United Kingdom\\
% ...
$^{78}$Department of Physics and Astronomy, Uppsala University, Uppsala, Sweden, associated to $^{57}$\\
% ...
}
\end{flushleft}

\end{document}